\documentclass{aa}  

\usepackage{graphicx}
\usepackage{txfonts}
\usepackage{xspace}
\usepackage{color}
\usepackage{url}
\usepackage{hyperref}
\newcommand{\lum}{erg s$^{-1}$}
\newcommand{\flux}{erg~s$^{-1}$~cm$^{-2}$}
\newcommand{\Nu}{{\it NuSTAR\xspace}}
\newcommand{\Nustar}{{\it NuSTAR\xspace}}
\newcommand{\hx}{{\it Insight-HXMT\xspace}}
\newcommand{\source}{4U 1901+03\xspace}

\begin{document}

\title{Spectral evolution of X-ray pulsar 4U 1901$+$03 during the 2019 outburst based on \hx\ and \Nu\ observations}

\author{Armin Nabizadeh \inst{1} 
        \and Sergey S. Tsygankov \inst{1,2}
                  \and Long Ji \inst{3}
                  \and Victor Doroshenko \inst{3,2}
                  \and Sergey V. Molkov \inst{2}
                  \and Youli Tuo \inst{4}
                  \and Shuang-Nan Zhang\inst{4}
                  \and Fan-Jun Lu\inst{4}
                  \and Shu Zhang\inst{4}
          \and Juri Poutanen \inst{1,2,5}
          }
          
   \institute{Department of Physics and Astronomy, FI-20014 University of Turku,  Finland \\ \email{armin.nabizadeh@utu.fi}
       \and
       Space Research Institute of the Russian Academy of Sciences, Profsoyuznaya Str. 84/32, Moscow 117997, Russia    
       \and
       Institut f\"ur Astronomie und Astrophysik, Universit\"at T\"ubingen, Sand 1, D-72076 T\"ubingen, Germany
       \and 
       Key Laboratory of Particle Astrophysics, Institute of High Energy Physics, Chinese Academy of Sciences, Beijing 100049, China
       \and
       Nordita, KTH Royal Institute of Technology and Stockholm University, Roslagstullsbacken 23, SE-10691 Stockholm, Sweden
          }
          
\titlerunning{Spectral properties of 4U 1901+03}
\authorrunning{Nabizadeh et al.}

\date{2021}

\abstract{
We report on a detailed spectral analysis of emission from X-ray pulsar \source\ using data obtained by the \textit{Insight-HXMT} and \textit{NuSTAR} observatories during the 2019 outburst of the source. Thanks to the extensive coverage of the outburst by  \textit{Insight-HXMT}, we were able to investigate the spectral evolution of the source as a function of flux, and compare these results to the previous reports, focusing on the properties of a putative absorption feature at around 10~keV. In particular, we demonstrate that the broadband X-ray continuum of \source\ can be well described with a two-component continuum model without an absorption line at 10~keV, which casts doubt on its interpretation as a cyclotron line. The high quality of the data also allowed us to perform both phase-averaged and phase-resolved spectral analyses as a function of luminosity. Finally, we performed a detailed investigation of another absorption feature in the spectrum of the source around 30~keV recently reported in the \Nu\ data.
We show that this feature appears to be significantly detected both in phase-averaged and phase-resolved spectra irrespective of the continuum model.
}


\maketitle
%
\section{Introduction}
X-ray pulsars (XRPs) are rotating, strongly magnetized neutron stars (NSs) accreting matter from a non-degenerate companion star. The accreting material is funneled by the magnetic field from either an accretion disc or a strong stellar wind to the NS surface,  producing pulsed X-ray emission. The X-ray spectra of some of the known XRPs show absorption-like features, the so-called cyclotron resonant scattering features \citep[CRSFs; see e.g. ][]{Staubert2019}, which are caused by resonant scattering of photons by the electrons in the accretion stream. 
Such features, if detected, provide the only direct method to estimate the strength of the NS magnetic field.

\source\ is a high-mass X-ray binary (HXMB) which was first detected by \textit{Uhuru} and \textit{Vela 5B} during a giant outburst in 1970--1971 \citep{1976ApJ...206L..29F,1984ApJ...280..661P}. The source then remained in quiescence until February 2003 when it was detected again by the \textit{Rossi X-ray Timing Explorer}/All Sky Monitor (\textit{RXTE}/ASM) during another giant outburst. This detection triggered an extensive observational campaign  \citep{Galloway2003b,Galloway2005}, which allowed the source to be studied in a broad range of luminosities up to $\sim$10$^{38}$ \lum\ \citep[for the assumed distance of 10 kpc;][]{Galloway2005}, and pulsations were detected with a period of around 2.763\,s, clearly establishing the source as an XRP. The pulsar timing  also allowed us to determine the orbital parameters of the binary system: the orbital period $P_{\rm orb}=22.58$~d, the eccentricity $e=0.036$, and the projected semi-major axis  $a_{\rm x}\sin i=107$ lt-s \citep{Galloway2005}.

Although no optical counterpart has been identified, the source was classified as a transient Be/X-ray pulsar  based on its orbital parameters and observed outbursts  \citep{Galloway2005}. Using the optical and infrared (IR) observations, \citet{Reig2016} were unable to identify obvious candidates for the optical counterpart, largely because of the poor X-ray localisation. Later in 2019, \citet{2019ATel12514....1M} and \citet{Hemphill2019ATel} improved the source localisation significantly using the \textit{Swift}/XRT and \textit{Chandra}/HRC data. This allowed the authors to propose the IR/optical candidate star J190339.39+031215.6 in the UKIDS-DR6 catalogue as the optical counterpart to \source. The optical spectrum of the star was subsequently obtained and an early spectral type was confirmed, consistent with the Be classification \citep{Strader2019ATel, McCollum2019ATel}.

According to the \textit{Gaia} measurements, the distance to the source is 3.0$^{+2.0}_{-1.1}$ kpc \citep{BailerJones2018}. However, \citet{Strader2019ATel} argued that the parallax measurement of the optical counterpart in \textit{Gaia} DR2 is not significant and thus this estimate is not constraining. Furthermore, the PS1 reddening maps \citep{2018MNRAS.478..651G}, high X-ray absorption, and optical reddening in the direction of the source also point to a larger distance of >12 kpc \citep{Strader2019ATel}. A similarly large distance of $d=12.4\pm0.2$~kpc was recently derived by \citet{Tuo2020} based on the torque models and the pulse profile evolution during the 2019 outburst.

The pulse profile of \source\ is relatively complex, and demonstrates variable structure depending both on the X-ray luminosity and photon energy \citep{Chen2008,Lei2009}. Based on the \textit{NICER} and \hx\ data obtained during the 2019 outburst, \cite{LongJi2020} discovered short flares in the light curve of the source with significant variability of the pulse profile in the flares and persistent emission. The observed behaviour was interpreted as changes in the beam pattern due to transitions between sub- and supercritical accretion regimes and the onset of an accretion column.

Assuming the spin equilibrium, \citet{Galloway2005} estimated the magnetic field strength of the NS to be $B \approx$ 0.3 $\times$ 10$^{12}$($d/10$ kpc)$^{-6/7}$ G. Later, \citet{James2011} interpreted the discovered quasi-periodic oscillations (QPOs) at around 0.135 Hz in the context of the beat-frequency model \citep{Alpar-Shaham1985,Miller1998}. Assuming that the QPO appears at the beat frequency between the pulsar spin frequency and the Keplerian frequency
at the magnetospheric radius, the authors obtained the same value of $B \approx$ 0.3 $\times$ 10$^{12}$ G (assuming distance of 10 kpc). On the other hand, the spin evolution of \source\ during the 2019 outburst led \citet{Tuo2020} to argue in favour of ten-times stronger field $B \approx$ 4.3 $\times$ 10$^{12}$ G, leaving questions as to the NS magnetic field open.

\begin{figure}
\begin{center} 
\includegraphics[width=\columnwidth]{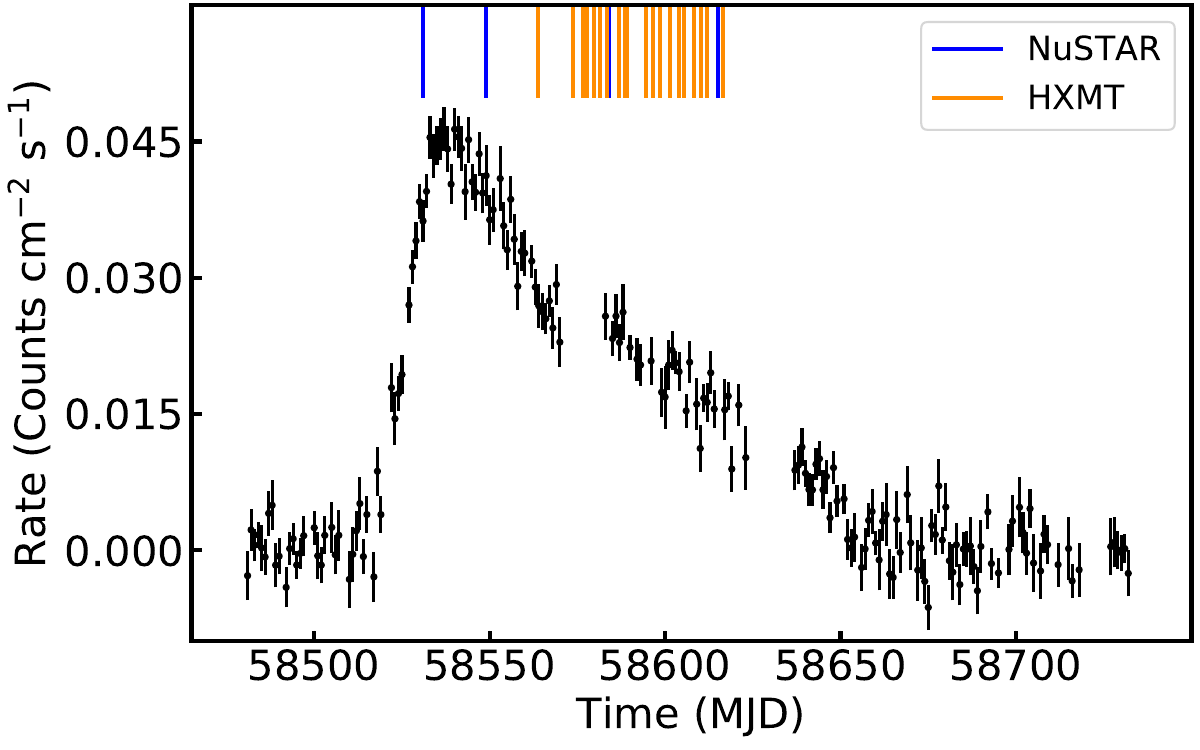}
\end{center}
\caption{15--50 keV X-ray light curve obtained by \textit{Swift}/BAT during the source outburst. The orange and blue vertical solid lines show the dates of \textit{HXMT} and \textit{NuSTAR} observations, respectively. 
}
\label{fig:bat} 
\end{figure}

The X-ray spectrum of \source\ has a shape similar to those of other accreting pulsars. \citet{Galloway2005} modelled the \textit{RXTE} data with a thermal Comptonization model. We note that although some residuals around 10 keV were still visible when applying this model \citep[see Fig. 2 in][]{Galloway2005}, the authors did not claim a detection of a cyclotron resonant scattering feature (CRSF) in the spectrum at this energy. Later, \citet{Reig2016} reanalysed the \textit{RXTE} observations concluding that it is not possible to fit the X-ray spectrum without a Gaussian absorption component that accounts for the absorption-like feature around 10 keV. The authors also found a positive correlation between the energy of the absorption feature and X-ray flux. This makes \source\ one of a handful of objects displaying such correlation \citep{Staubert2019}, and allowed \citet{Reig2016} to tentatively suggest that the 10-keV feature is indeed associated with a cyclotron line. This also implies that \source\ remained in the subcritical accretion regime during the 2003 outburst.  
In addition, \citet{Coley2019ATel} recently reported negative residuals from their best-fit model around 30~keV using \Nu\ observations during the declining phase of the 2019 outburst, suggesting that it could be a cyclotron absorption line.

Here we report a detailed broad-band spectral analysis of \source\ in a wide range of X-ray luminosities using independent high-quality data obtained by the \Nu\ and \hx\ observatories during the 2019 outburst. Specifically, we aim to clarify the origin of both the 10- and 30 keV absorption features.


\begin{table}
    \centering
    \caption{Observation log of \source\ during its 2019 outburst with \hx\ and \Nu.}
    \begin{tabular}{cccc}
    \hline\hline
    ObsID & Start date & Start MJD & Exposure (ks) \\
    \hline
    \multicolumn{4}{c}{\hx}  \\
    P0211006001 & 2019-03-21 & 58563.81 & 5.35 \\
    P0211006003 & 2019-03-31 & 58573.85 & 3.31 \\
    P0211006004 & 2019-04-03 & 58576.59 & 1.53 \\
    P0211006005 & 2019-04-04 & 58577.77 & 1.89 \\
    P0211006006 & 2019-04-06 & 58579.76 & 2.26 \\
    P0211006007 & 2019-04-08 & 58581.34 & 2.11 \\
    P0211006008 & 2019-04-10 & 58583.33 & 2.22 \\
    P0211006009 & 2019-04-13 & 58586.87 & 2.03 \\
    P0211006010 & 2019-04-15 & 58588.54 & 2.75 \\
    P0211006011 & 2019-04-16 & 58589.20 & 1.95 \\
    P0211006012 & 2019-04-21 & 58594.59 & 1.83 \\
    P0211006013 & 2019-04-23 & 58596.43 & 2.25 \\
    P0211006014 & 2019-04-25 & 58598.52 & 3.38 \\
    P0211006016 & 2019-04-28 & 58601.32 & 3.09 \\
    P0211006017 & 2019-04-30 & 58604.04 & 4.37 \\
    P0211006018 & 2019-05-02 & 58605.30 & 1.49 \\
    P0211006019 & 2019-05-04 & 58608.14 & 5.28 \\
    P0211006020 & 2019-05-06 & 58610.12 & 4.88 \\
    P0211006021 & 2019-05-08 & 58612.05 & 4.70 \\
    P0211006022 & 2019-05-12 & 58616.46 & 10.00 \\
    \multicolumn{4}{c}{\Nu}  \\
    90501305001 & 2019-02-17 & 58531.12 & 17.85 \\
    90502307002 & 2019-03-07 & 58549.31 & 12.25 \\
    90502307004 & 2019-04-11 & 58584.94 & 21.45 \\
    90501324002 & 2019-05-12 & 58615.74 & 45.12 \\

    \hline
    \end{tabular}
    \label{tab:observations}
\end{table}

\section{Observations and data reduction}
\label{sec:observations}

\begin{figure*}
\begin{center} 
\includegraphics[width=0.9\columnwidth]{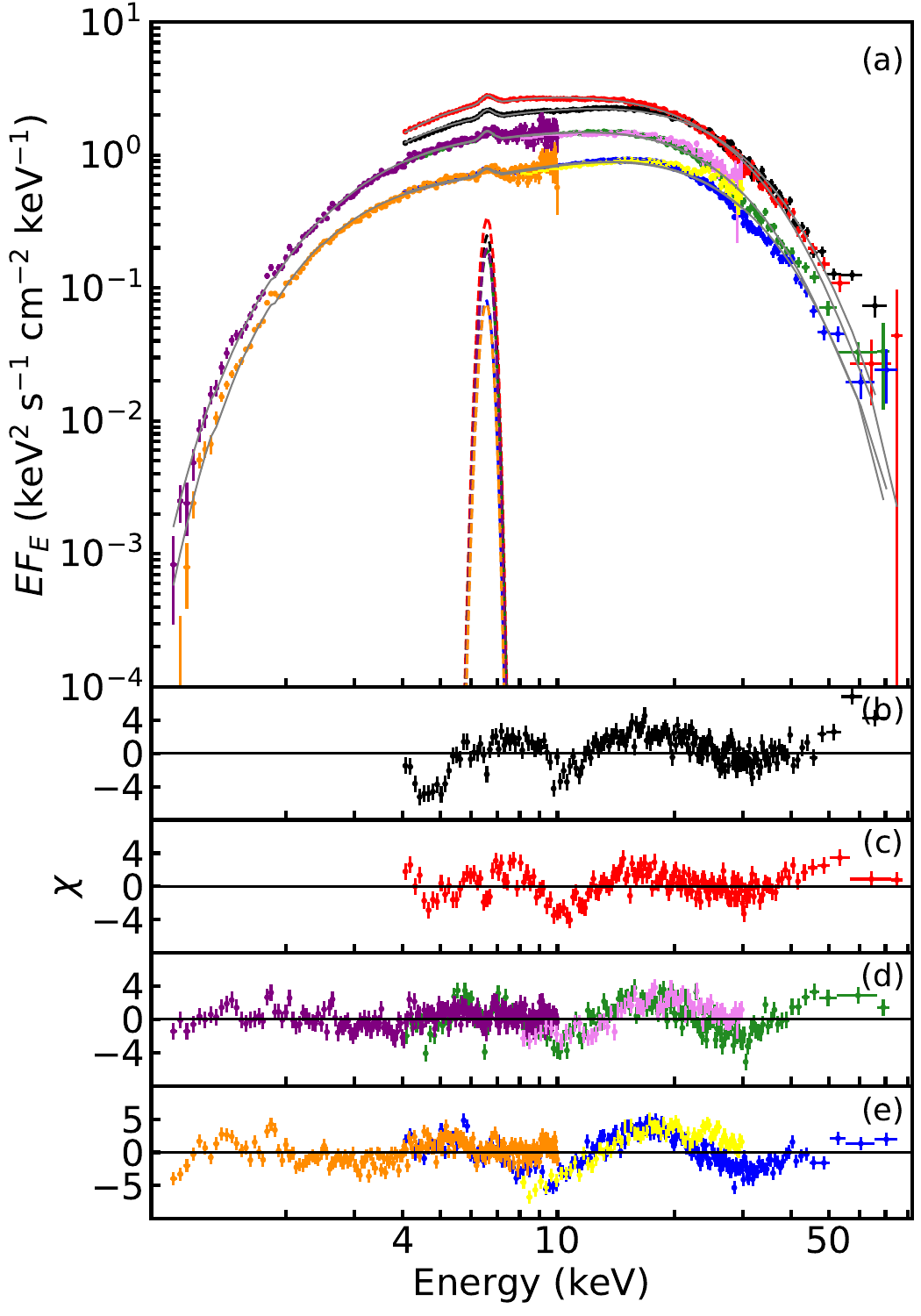}\hspace{1cm}
\includegraphics[width=0.9\columnwidth]{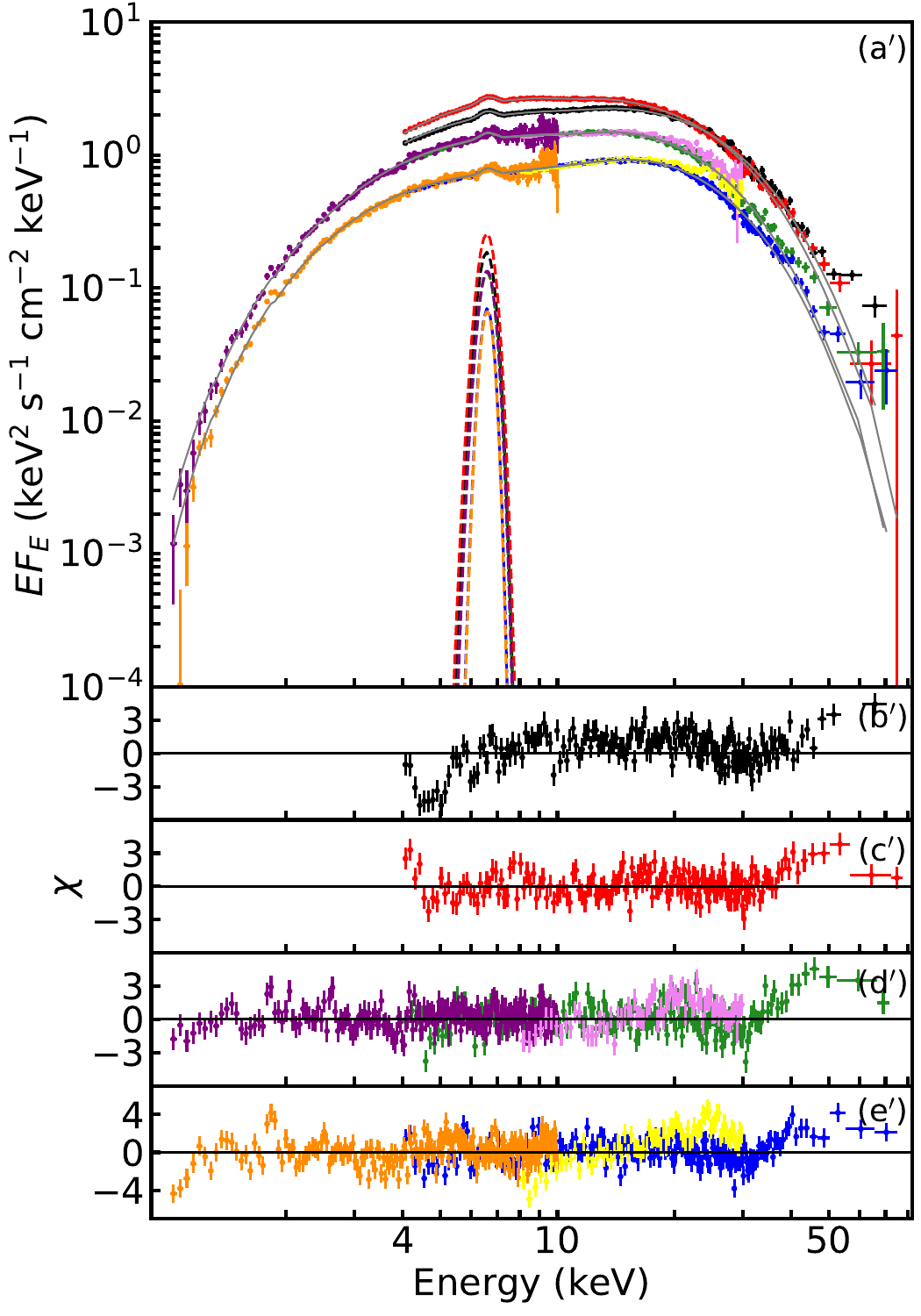}
\end{center}
\caption{\textit{Left panels:} 
X-ray spectra of \source\ obtained by \hx\ (LE and ME detectors in purple and pink for third data set and orange and yellow for fourth data set, respectively) and \Nu\ (from the first to the last chronologically in black, red, green, and blue, respectively. Only FPMA data are shown here for clarity.) during the outburst together with the composite model {\sc tbabs $\times$ (comptt + gaussian)} (solid grey curves) shown in \textit{panel a}. The dashed curves with the same colours show the iron line component.
Residuals for the four data sets using the corresponding composite model are shown in units of standard deviations \textit{(panels b--e)}. 
\textit{Right panels:}  Same X-ray spectra as in \textit{panel a} fitted with the same composite model but with a {\sc gabs} component added to account for the absorption feature at 10 keV (\textit{panel a$^\prime$}). Residuals of the same spectra using the corresponding composite model are shown in \textit{panels b$^\prime$--e$^{\,\prime}$}.
}
\label{fig:nustar-fit} 
\end{figure*}

The 2019 outburst of the source has been extensively monitored by several instruments. Here we focus on the analysis of data obtained by the \Nu\ and \hx\ observatories both of which cover a broad range of energies essential for spectral analysis. A summary of the observations is provided in Table~\ref{tab:observations} and Fig.~\ref{fig:bat}. The latter shows the \textit{Swift}/BAT long-term light curve with the dates of the \hx\ and \Nu\ observations indicated. Details of the data reduction for both instruments are presented below.

\subsection{\hx\ observations}

Hard X-ray Modulation Telescope (\textit{HXMT}; also known as \hx), is China's first X-ray telescope and was launched on 2017 June 15 \citep{LI2007131,Zhang2020}. The telescope has a large effective area in hard X-rays, and covers a broad energy band 1--250 keV. The three slat-collimated instruments onboard are the low-energy X-ray telescope (LE), the medium-energy X-ray telescope (ME), and the  high-energy X-ray telescope (HE) operating in the energy ranges  1--15, 5--30, and 20--250 keV, respectively \citep{Zhang2020,Liu2020,Cao2020,Chen2020}. 

During the declining phase of the outburst,  \source\ was observed 20 times by \hx\  over the period from MJD 58563 to MJD 58616. This long duration monitoring provided us with a good opportunity to study the spectral evolution of the source at different luminosities. We note that the background starts to dominate at energies above $\sim$50~keV even close to the peak of the outburst, and so the HE data unfortunately provide no useful information. Therefore, for the present study we decided to use only LE and ME data covering the energy range between 1 and 30 keV. In order to reduce the HXMT data and extract the source spectra, we performed the standard data-reduction procedure explained in the \hx\ official user guide\footnote{\url{http://enghxmt.ihep.ac.cn/sjfxwd/168.jhtml}} using the software {\scriptsize HXMTSOFT} v2.01 with a {\sc caldb} v2.01 \citep{Li2020, Liao2020, Liao2020b, Guo2020}.

\begin{figure}
\begin{center} 
\includegraphics[width=0.9\columnwidth]{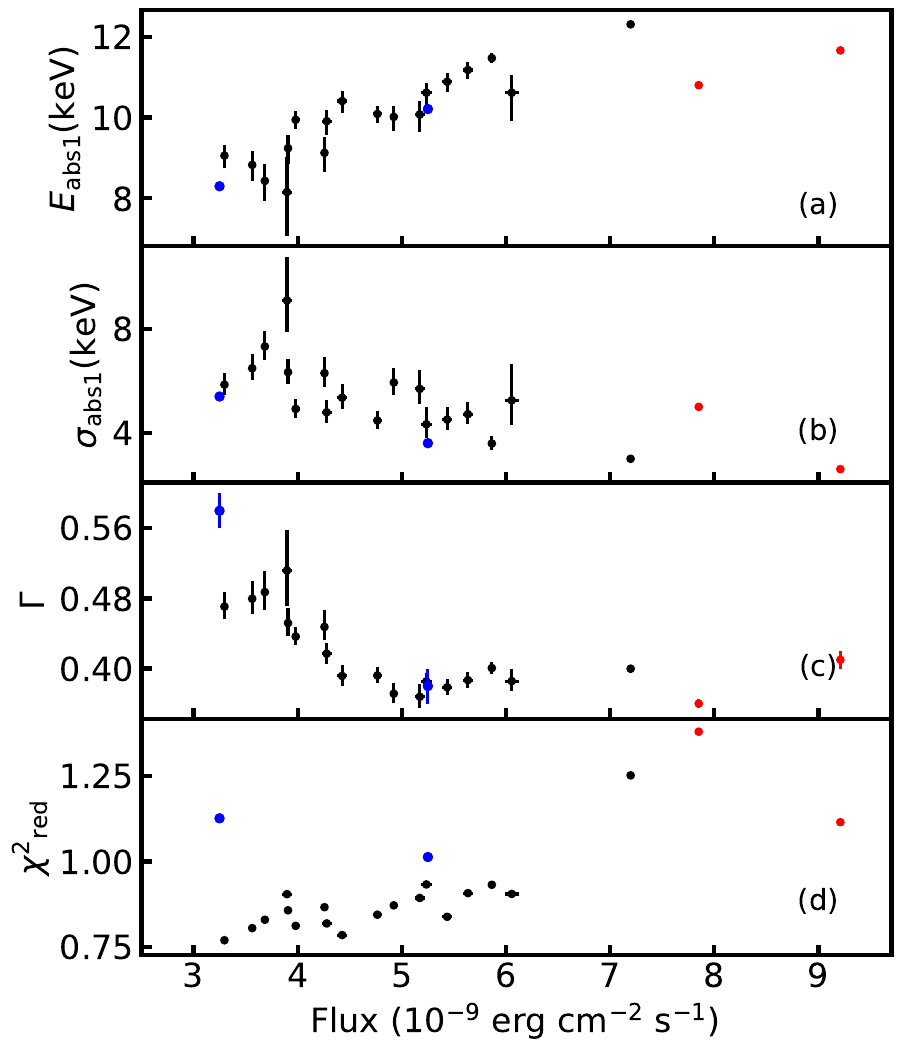}
\end{center}
\caption{Evolution of the 10-keV absorption feature energy $E_{\rm abs1}$ \textit{(panel a)}, its width $\sigma_{\rm abs1}$ \textit{(panel b)}, the photon index $\Gamma$ \textit{(panel c)}, and the reduced $\chi^2$ of the fits \textit{(panel d)} as a function of the 1--30 keV flux obtained from the \hx\ data (black points). Red and blue dots show the same parameters derived from two single \Nu\ observations and two combined observations (\Nu+\hx), respectively.  We used the model {\sc tbabs $\times$ (gau+cutoffpl $\times$ gabs)} here with the $N_{\rm H}$ and $E_{\rm cut}$ fixed at 2.4$\times$10$^{22}$ cm$^{-2}$ and 6.66 keV, respectively (see the text).
}
\label{fig:reig} 
\end{figure}

\subsection{\Nu\ observations}
\label{nu-observations}

The Nuclear Spectroscopic Telescope Array (\Nu) is the first focusing hard X-ray observatory and is equipped with two identical co-aligned X-ray telescopes, Focal Plane Module A and B (FPMA and FPMB) \citep{Harrison2013}. The instruments cover a wide energy range 3--79 keV and provide an imaging resolution of 18\arcsec\ (full width at half maximum, FWHM) and a spectral energy resolution of 400 eV (FWHM) at 10 keV. \source\ was observed by \Nu\ four times during its recent outburst in a four-month period from February to May 2019 (see Table~\ref{tab:observations} for more details). 
The first observation (ObsID 90501305001) was performed at the rising phase of the outburst while the others were done at the declining phase (see Fig.~\ref{fig:bat}). The uniform distribution of the observations over the outburst provides us with a good opportunity to study the evolution of the source properties as a function of luminosity and compare these results to the ones obtained by \hx. 
From now on, we refer to the four \Nu\ observations as NuObs1, NuObs2, NuObs3, and NuObs4, named in the chronological order.

In order to reduce the raw data and extract the source light curves and spectra, we performed the standard data reduction procedure following the \Nu\ official user guides.\footnote{\url{https://nustar.ssdc.asi.it/news.php\#}} 
We used the \Nu\ Data Analysis Software {\sc nustardas} v2.0.0 with a {\sc caldb} version 20201130. A source-centred circular region with a radius of 120\arcsec\ was chosen to extract the source photons from both FPMA and FPMB. The background was extracted similarly from a source-free circular region of the same size. The spectra collected from \Nu/FPMA and FMPB were simultaneously fitted in {\sc xspec} in the energy range 4--79 keV. In addition, some of the \hx\ data sets are performed quasi-simultaneously with two \Nu\ observations: \hx\ P0211006008 with NuObs3 and \hx\ P0211006022 with NuObs4. The joint temporal and spectral studies on Her X-1 using data from \hx\ and \Nu\ showed that the two observatories revealed highly consistent results \citep{NuSTAR-HXMT-Calibration2019}, indicating no calibration problems. Therefore, to find the spectral model that best describes the broadband continuum spectrum of the source, 
we define four data sets as NuObs1 (\#1), NuObs2 (\#2), NuObs3+P0211006008 (\#3), and NuObs4+P0211006022 (\#4) to be used for the detailed spectroscopy.

In order to use $\chi^2$ statistics, the \hx\ and \Nu\ spectra were each grouped to have at least 30 and 25 counts in each energy bin, respectively, for phased-average and phased-resolved spectral analyses. For all \hx\ and \Nu\ observations, we used {\sc heasoft} 6.28\footnote{\url{http://heasarc.nasa.gov/lheasoft/}} and {\sc xspec} 12.11.0m\footnote{\url{https://heasarc.gsfc.nasa.gov/xanadu/xspec/manual/XspecManual.html}} to perform the spectral analysis.

The pulse phase-resolved spectral analysis also requires determination of the timing properties of the source. To obtain the pulse profile of each observation, we extracted the light curves from the same regions and then applied the barycentric correction to them using the standard {\sc barycorr} task (see Sect.~\ref{phase-resolved}).

\begin{figure}
\begin{center} 
\includegraphics[width=0.9\columnwidth]{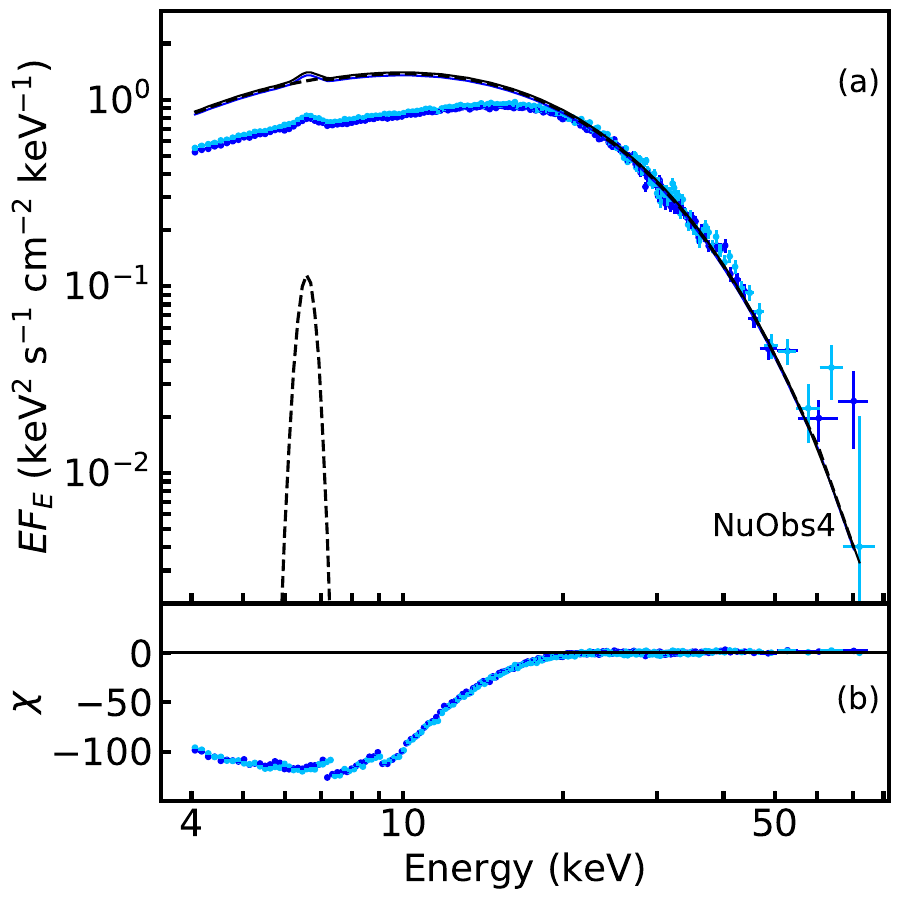}
\end{center}
\caption{{\it Panel a}: Source spectrum obtained during NuObs4 together with the model {\sc tbabs $\times$ (gau+cutoffpl $\times$ gabs)} where the absorption line depth was set to zero (solid line). {\it Panel b}: The corresponding residuals. 
}
\label{fig:gabs-anomoly} 
\end{figure}

\begin{figure}
\begin{center} 
\includegraphics[width=0.9\columnwidth]{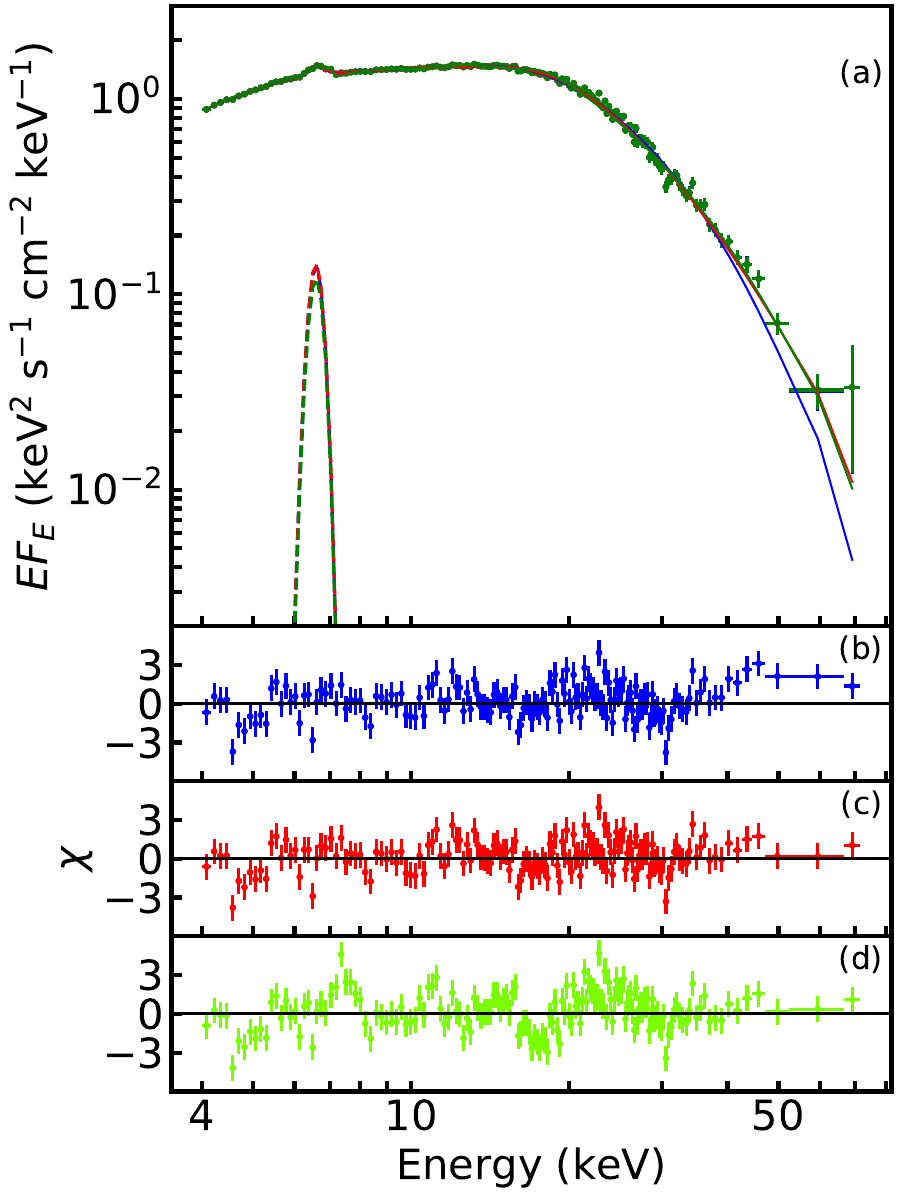}
\end{center}
\caption{\textit{Panel a}: X-ray spectrum of \source\ taken from NuObs3 together with the composite model {\sc tbabs $\times$  (gau + continuum $\times$ gabs)} (solid lines) where {\sc continuum} stands for  {\sc cutoffpl}, {\sc npex}  and  {\sc po $\times$ highecut} and shown by blue, red and green curves, respectively. The corresponding residuals are shown in \textit{panels b, c} and \textit{d}.  
Only FPMA data are shown for the sake of clarity.}
\label{fig:test-fits} 
\end{figure}

\section{Results}
\subsection{Pulse phase-averaged spectroscopy}
\label{Phase-averaged}

Similarly to other accreting XRPs, the spectrum of \source\ has a smooth power-law-like shape with a cutoff at high energies \citep[see e.g.][]{Filippova2005}. 
We attempted to use different physical and phenomenological continuum models such as: a power law with a high-energy cutoff model {\sc cutoffpl} described by the equation $N(E) = K{E^{-{\rm \alpha}}} {\rm exp}(-E/\beta)$; a Comptonization model {\sc comptt} \citep{Titarchuk1994}; a power law with a high-energy cutoff model {\sc po $\times$ highecut}, with the latter defined by the equation $M(E)=\exp[(E_{\rm c} - E)/E_{\rm f}]$ for ($E \geq E_{\rm c}$) and $M(E) = 1$ for ($E \leq E_{\rm c}$); a {\sc npex} model, representing the sum of power laws of negative and positive slopes given by equation $N(E) = (A_{\rm 1}E^{-\alpha_{\rm 1}} + A_{\rm 2}E^{+\alpha_{\rm 2}})~{\rm exp}(-E/kT)$ \citep{Mihara1998}; and a power-law model with the Fermi-Dirac cutoff {\sc fdcut} described by equation $N(E) = A_{\rm PL} {E^{-\Gamma}}/[{{\rm exp}((E - E_{\rm cut})/E_{\rm fold}) + 1}]$  \citep{Tanaka1986}. We also included a Gaussian emission component {\sc gau} corresponding to the fluorescent iron line at 6.4~keV and modified the continuum by the photoelectric  absorption to account for interstellar and intrinsic absorption {\sc tbabs} \citep{Wilms2000} with default abundances from \citet{An-Gre1989} and atomic cross-sections adopted from \citet{Verner1996}. A cross-normalisation constant was also included to account for minor differences in the absolute calibration of the effective areas of the individual instruments. We note that none of the mentioned models used in this analysis give a physical description of the X-ray emission from XRPs as the spectral formation is much more complex in these systems. In recent years, several physically motivated models have been proposed by different authors \citep[see e.g. ][]{2007ApJ...654..435B,2012A&A...538A..67F,2017ApJ...835..130W,2021MNRAS.501..564G}; however, applying such models to fit the spectrum of \source is beyond the scope of this study.

Application of the models listed above lead to  fits of unacceptable quality. 
For example, for the data set \#3 we obtain $\chi^2$ values of 6781 (for 2957 d.o.f.), 3927 (2956), 5808 (2955), 3720 (2954), and 4425 (2956), respectively, for {\sc cutoffpl}, {\sc comptt}, {\sc po $\times$ highecut}, {\sc npex,} and  {\sc po $\times$ fdcut}.  The fits produced residuals around 10 and 30 keV (see Fig.~\ref{fig:nustar-fit}-left). Therefore, the source spectrum is more complex than predicted by any of the  above-mentioned models. A detailed description of the fitting results for individual models is provided below.

\begin{figure}
\begin{center} 
\includegraphics[width=0.9\columnwidth]{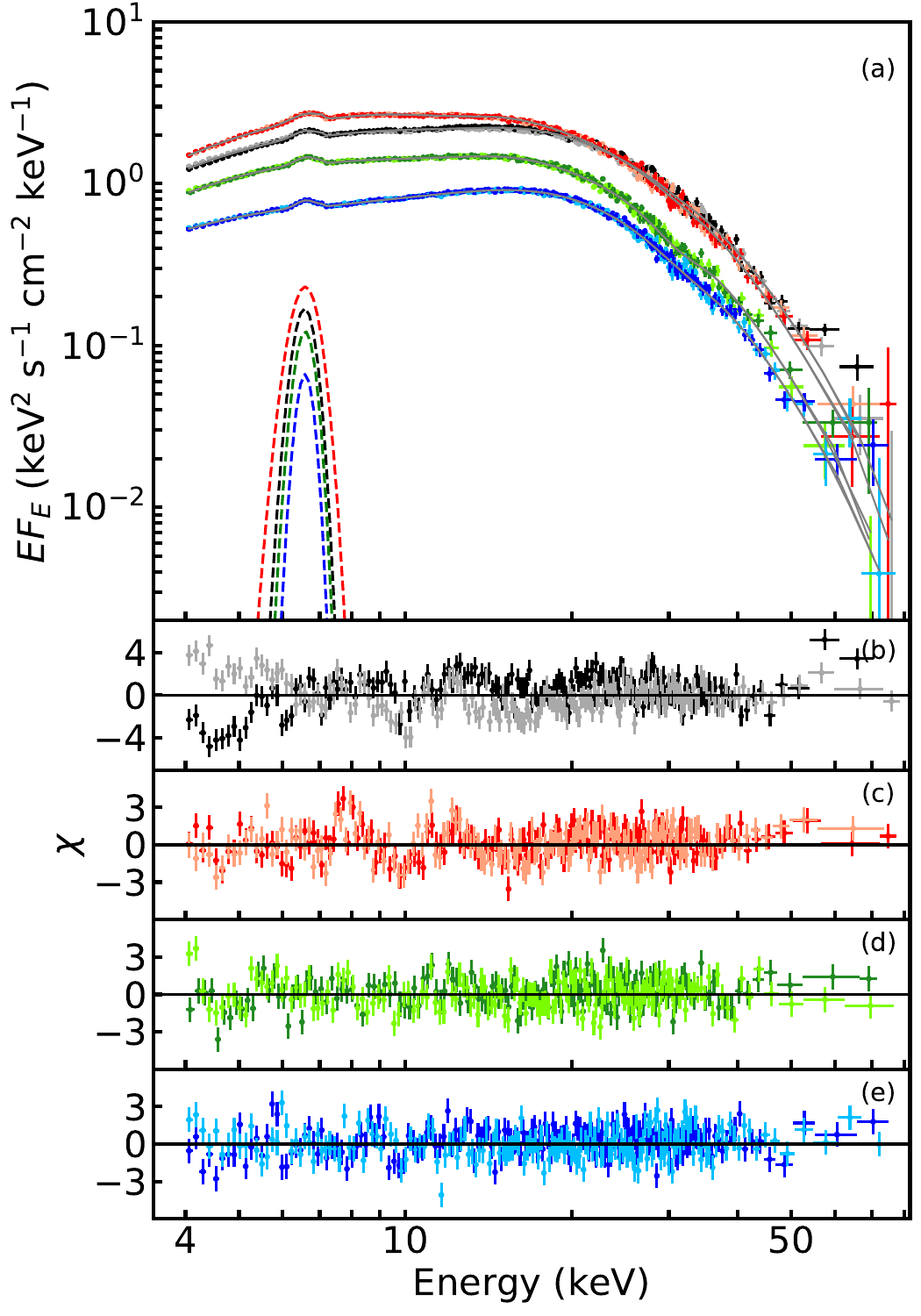}
\end{center}
\caption{Same as Fig.~\ref{fig:nustar-fit}, but only for the \Nu\ observations together with the best-fit model ({\sc tbabs $\times$ (gau+comptt $\times$ gabs $\times$ gabs))}. Lighter colours represent the data obtained from the  \Nu/FPMB module.
}
\label{fig:best-fits1} 
\end{figure}
\begin{figure}
\begin{center} 
\includegraphics[width=0.9\columnwidth]{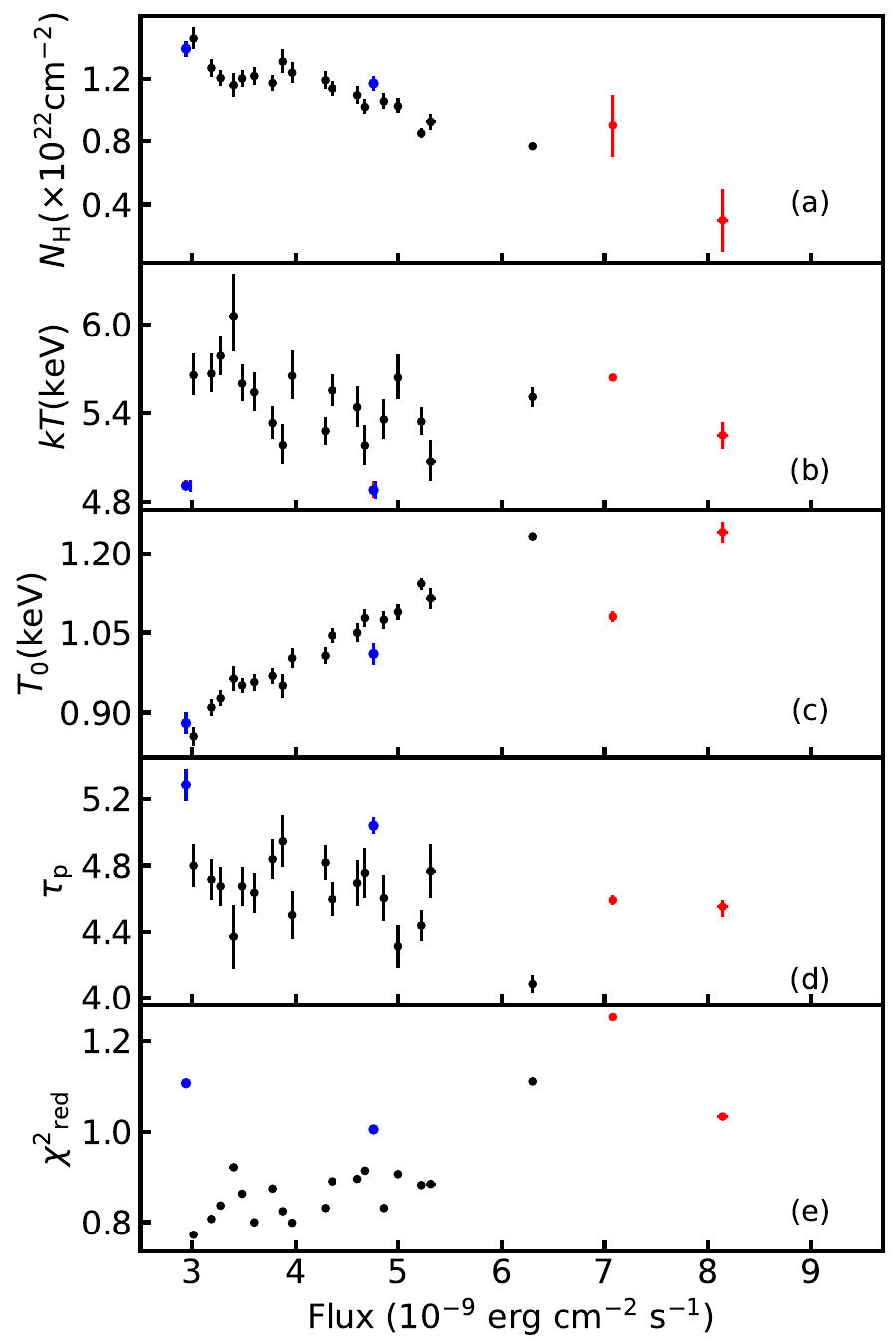}
\end{center}
\caption{Evolution of the interstellar absorption $N_{\rm H}$ \textit{(panel a)}, the plasma temperature $kT$ \textit{(panel b)}, the input photon temperature $T_{\rm 0}$ \textit{(panel c)}, the plasma optical depth  $\tau_{\rm p}$ \textit{(panel d)}, and the reduced $\chi^2$ of the fits \textit{(panel e)} as a function of the 1--30 keV flux obtained from the \hx\ spectra using the {\sc tbabs $\times$ (gau+cutoffpl $\times$ gabs)} model. Red and blue dots show the same parameters derived from two single \Nu\ observations and two combined observations (\Nu+\hx), respectively. 
}
\label{fig:reig-comptt} 
\end{figure}

\subsubsection{Continuum model {\sc cutoffpl}}

To be able to make a comparison between our results and previous findings from the literature, in the first step we adopted the model of a power law with a high-energy exponential cutoff ({\sc cutoffpl} in {\sc xspec}) with the additional Gaussian absorption component {\sc gabs} to fit the absorption structure around 10 keV.
Such a model was used by \citet{Reig2016} to fit the \textit{RXTE} spectra of the source. 
The composite model {\sc tbabs $\times$ (gau+cutoffpl $\times$ gabs)} revealed acceptable fits with $\chi^2$(d.o.f.) values of 2102 (1668), 1720 (1568), 3081 (2954), and 3522 (3090), respectively, for our four data sets. The fit parameters and the corresponding uncertainties are given in Table~\ref{tab:2b}. All the uncertainties in the spectral analysis are given at 68.3\% (1$\sigma$) confidence level everywhere in this paper. We also applied this model to all 20 \hx/LE+ME spectra to study the evolution of spectral parameters as a function of X-ray flux in the energy range 1--30 keV. Of particular interest is the dependence of the putative 10~keV absorption feature centroid energy on flux. However, when the results from two telescopes ---where all parameters can be fitted independently--- are compared,  we find that there are inconsistencies between the values of the parameters $E_{\rm abs1}$, $\Gamma$, $N_{\rm H}$, and $E_{\rm cut}$ most probably caused by the different composition of the data sets, that is, a continuous broadband spectrum for \Nustar\ in contrast to two joint bands in \hx\ and different energy bands covered by the instruments. As a result, the cross-normalisation constant used in the \hx\ data adds one additional free parameter.

In order to obtain more consistent results, we  used  two combined spectra (\Nustar\ + \hx), corresponding to data set \#3 and data set \#4, and found the values of parameters $N_{\rm H}$ and $E_{\rm cut}$ , which cannot be constrained by the fit using the \hx\ data alone. Therefore, for the NuObs1 and NuObs2 and the remaining 18 \hx\ observations, we fixed the $N_{\rm H}$ and $E_{\rm cut}$ to values of 2.4 $\times$ 10$^{22}$ cm$^{-2}$ and 6.66 keV, respectively, which were obtained by averaging the corresponding best-fit values for data sets \#3 and \#4 to avoid this inconsistency. As shown in Fig.~\ref{fig:reig}, the energy of the 10-keV feature appears to be correlated with X-ray flux, which is in full agreement with the \textit{RXTE} results \citep{Reig2016}. The photon index also shows some evolution, decreasing with increasing flux, similarly to the \textit{RXTE} data.

\begin{table*}
\begin{center}
        \caption{Best-fit parameters for the model {\sc tbabs $\times$ (gau+cutoffpl $\times$ gabs)} for the four data sets of pulse-averaged spectra during the 2019 outburst.}
        \label{tab:2b}
        \begin{tabular}{lcccccc} 
        \hline
        \hline
                Model & Parameters & Units                                  & NuObs1               & NuObs2                & NuObs3                     & NuObs4 \\
                                                 &                  &       &                      &                       & +P0211006008               & +P0211006022\\
        \hline
        \textsc{constant}\tablefootmark{a}   &      FPMB/FPMA        &   &  0.996$^{+0.001}_{-0.001}$     & 1.026$\pm$0.001       & 1.028$^{+0.001}_{-0.001}$   & 1.017$^{+0.001}_{-0.001}$ \\
        \textsc{constant}\tablefootmark{b}   &      LE/FPMA      &   &                                &                       & 0.987$^{+0.004}_{-0.004}$   & 0.930$^{+0.003}_{-0.003}$ \\
        \textsc{constant}\tablefootmark{c}   &      ME/FPMA       &   &                                &                       & 1.004$^{+0.003}_{-0.003}$   & 0.983$^{+0.002}_{-0.002}$ \\

    \textsc{tbabs}  &  $N_{\rm H}$    & $10^{22}$ cm$^{-2}$     &  2.3$^{+0.2}_{-0.2}$                 & 2.9$^{+0.2}_{-0.2}$             & 2.31$^{+0.04}_{-0.04}$       & 2.57$^{+0.04}_{-0.04}$ \\

        \textsc{cutoffpl}   &  $\Gamma$            &                &  0.50$^{+0.01}_{-0.01}$   & 0.52$^{+0.02}_{-0.02}$    & 0.38$^{+0.01}_{-0.01}$    & 0.58$^{+0.02}_{-0.02}$ \\
                        &  $E_{\rm cut}$       & keV            &  7.42$^{+0.05}_{-0.05}$   & 7.00$^{+0.06}_{-0.06}$    & 6.39$^{+0.05}_{-0.05}$       & 6.94$^{+0.05}_{-0.05}$ \\
                                    &  $A_{\rm PL}$        &                &  0.314$^{+0.003}_{-0.003}$ & 0.40$^{+0.01}_{-0.01}$    & 0.211$^{+0.003}_{-0.003}$    & 0.217$^{+0.01}_{-0.009}$ \\

        \textsc{gabs}   &  $E_{\rm abs1}$          & keV                &  10.95$^{+0.05}_{-0.05}$  & 11.55$^{+0.06}_{-0.06}$   & 10.21$^{+0.04}_{-0.04}$   & 8.3$^{+0.1}_{-0.1}$ \\
                                    &  $\sigma_{\rm abs1}$     & keV            &  3.6$^{+0.1}_{-0.1}$      & 2.2$^{+0.1}_{-0.1}$       & 3.6$^{+0.1}_{-0.1}$       & 5.4$^{+0.1}_{-0.1}$ \\
                    &  $\tau_{\rm abs1}$          &                 &  0.17$^{+0.01}_{-0.01}$   & 0.10$^{+0.01}_{-0.01}$    & 0.23$^{+0.02}_{-0.02}$    & 0.49$^{+0.04}_{-0.04}$ \\

    \textsc{gaussian}   &  $E_{\rm Fe}$        & keV            &  6.55$^{+0.01}_{-0.01}$   & 6.54$^{+0.02}_{-0.02}$    & 6.58$^{+0.01}_{-0.01}$    & 6.59$^{+0.01}_{-0.01}$ \\
                                        &  $\sigma_{\rm Fe}$   & keV            &  0.30$^{+0.02}_{-0.02}$   & 0.38$^{+0.03}_{-0.02}$    & 0.26$^{+0.02}_{-0.02}$    & 0.17$^{+0.02}_{-0.02}$ \\
            &  $A_{\rm Fe}$ & $10^{-3}$ ph s$^{-1}$cm$^{-2}$    &  3.3$^{+0.2}_{-0.2}$     & 5.38$^{+0.3}_{-0.3}$        & 2.1$^{+0.1}_{-0.1}$       & 1.13$^{+0.07}_{-0.08}$ \\
        \hline
     $F_{\rm 1-79\,keV}$\tablefootmark{d}       & & $10^{-9}$ erg s$^{-1}$cm$^{-2}$ & 8.26$^{+0.04}_{-0.04}$   & 9.78$^{+0.06}_{-0.06}$ & 5.41$^{+0.01}_{-0.01}$ & 3.40$^{+0.01}_{-0.01}$ \\
     $F_{\rm 4-79\,keV}$\tablefootmark{d}       & & $10^{-9}$ erg s$^{-1}$cm$^{-2}$ & 6.62$^{+0.02}_{-0.02}$ & 7.70$^{+0.02}_{-0.02}$ & 4.270$^{+0.006}_{-0.006}$ & 2.610$^{+0.003}_{-0.003}$ \\
                 $\chi^2$(d.o.f.)  &            &                           & 2102 (1668)                    & 1720 (1568)           & 3081 (2954)                & 3522 (3090)\\ 
    \hline
                 
        \end{tabular}
\end{center}
\tablefoot{
\tablefoottext{a}{Cross-normalisation constant between FPMB and FMPA instruments on board \Nu.}
\tablefoottext{b}{Cross-normalisation constant between  \hx/LE and \Nu/FPMA instruments.}
\tablefoottext{c}{Cross-normalisation constant between \hx/ME  and  \Nu/FPMA instruments.}
\tablefoottext{d}{Unabsorbed X-ray flux.}
}
\end{table*}

As we see from Fig.~\ref{fig:reig}, the 10 keV absorption feature centroid energy correlates with the flux from the source. At the same time, the width of this feature demonstrates the opposite behaviour. In the lowest available data set (NuObs4+P0211006022), the line energy and width reach 8.3 and 5.4 keV, respectively. Similar results were obtained from the \textit{RXTE} observations of the source in which the 10 keV absorption feature centroid energy (width) dropped (rised) remarkably from $E\sim$11.8 ($\sigma\sim$ 2.5) keV at flux $\sim$8 $\times$ 10$^{-9}$ \flux\ to $E\sim$6.3 ($\sigma\sim$ 5) keV at flux $\sim$6 $\times$ 10$^{-10}$ \flux\ \citep[see Fig. 7 in  ][]{Reig2016}. Obviously, in such a situation, the absorption component cannot be considered to be a narrow feature but rather acts as a continuum spectral component. To illustrate the problem, in Fig.~\ref{fig:gabs-anomoly} we show the \source\ spectrum from NuObs4 with the 10-keV feature depth ($D_{\rm abs1}$) set to zero. This casts very strong doubt on the interpretation of this feature as a CRSF and/or the use of the continuum model {\sc cutoffpl} proposed by \cite{Reig2016}. Therefore, for further analysis we decided to use the broadly accepted Comptonization model {\sc comptt} instead of {\sc cutoffpl} to fit the source spectra.

\begin{table*}
\begin{center}
        \caption{Best-fit parameters for the model {\sc tbabs $\times$ (gau+comptt $\times$ gabs $\times$ gabs)}  for the four data sets of pulse-averaged spectra during the 2019 outburst.}
        \label{tab:2}
        \begin{tabular}{lcccccc} 
        \hline
        \hline
                Model & Parameters & Units                                  & NuObs1               & NuObs2                & NuObs3                     & NuObs4 \\
                                                 &                  &       &                      &                       & +P0211006008               & +P0211006022\\
        \hline
        \textsc{constant}\tablefootmark{a}   &      FPMB/FPMA        &   &  0.996$^{+0.001}_{-0.001}$     & 1.026$^{+0.001}_{-0.001}$       & 1.028$^{+0.001}_{-0.001}$   & 1.017$^{+0.001}_{-0.001}$ \\
        \textsc{constant}\tablefootmark{b}   &      LE/FPMA      &   &                                &                       & 0.995$^{+0.004}_{-0.004}$   & 0.935$^{+0.003}_{-0.003}$ \\
        \textsc{constant}\tablefootmark{c}   &      ME/FPMA      &   &                                &                       & 1.004$^{+0.003}_{-0.003}$   & 0.983$^{+0.002}_{-0.002}$ \\

    \textsc{tbabs}  &  $N_{\rm H}$    & $10^{22}$ cm$^{-2}$     &  0.9$^{+0.2}_{-0.2}$      & 0.3$^{+0.2}_{-0.2}$          & 1.17$^{+0.05}_{-0.05}$       & 1.42$^{+0.05}_{-0.05}$ \\

        \textsc{comptt}   &  $T_{\rm 0}$                   & keV           &  1.08$^{+0.01}_{-0.01}$   & 1.24$^{+0.02}_{-0.02}$    & 1.01$^{+0.01}_{-0.01}$       & 0.88$^{+0.02}_{-0.01}$ \\
                      &  $kT$                   & keV           &  5.64$^{+0.06}_{-0.06}$   & 5.25$^{+0.08}_{-0.09}$    & 4.88$^{+0.06}_{-0.05}$       & 4.91$^{+0.04}_{-0.04}$ \\
                      &  $\tau_{\rm p}$                 &               &  4.592$^{+0.04}_{-0.04}$   & 4.55$^{+0.04}_{-0.06}$    & 5.04$^{+0.05}_{-0.05}$       & 5.29$^{+0.08}_{-0.1}$ \\
                                  &  $A_{\rm comptt}$        &              &  0.148$^{+0.002}_{-0.002}$ & 0.186$^{+0.002}_{-0.003}$ & 0.122$^{+0.002}_{-0.002}$   & 0.080$^{+0.004}_{-0.002}$ \\

        \textsc{gabs}   &  $E_{\rm abs1}$          & keV                &  10.6$^{+0.1}_{-0.1}$     & 10.8$^{+0.2}_{-0.2}$     & 10.5$^{+0.1}_{-0.1}$       & 9.0$^{+0.4}_{-0.8}$ \\
                                    &  $\sigma_{\rm abs1}$     & keV            &  1.1$^{+0.2}_{-0.2}$      & 1.7$^{+0.2}_{-0.3}$      & 2.1$^{+0.2}_{-0.2}$        & 4.4$^{+0.8}_{-0.5}$ \\
                    &  $\tau_{\rm abs1}$          &                 &  0.03$^{+0.01}_{-0.01}$   & 0.05$^{+0.01}_{-0.02}$   & 0.06$^{+0.01}_{-0.01}$      & 0.14$^{+0.1}_{-0.05}$ \\

        \textsc{gabs}   &  $E_{\rm abs2}$          & keV                &  39.2$^{+0.4}_{-0.4}$     & 35.3$^{+0.4}_{-1.4}$     & 32.2$^{+0.6}_{-0.6}$      & 32.4$^{+0.6}_{-0.5}$ \\
                                    &  $\sigma_{\rm abs2}$     & keV            &  10.0$^{+3.4}_{-0.5}$     & 9.9$^{+2.0}_{-1.5}$      & 5.8$^{+0.6}_{-0.5}$       & 5.5$^{+0.8}_{-0.6}$ \\
                    &  $\tau_{\rm abs2}$          &                 &  0.46$^{+0.19}_{-0.06}$   & 0.37$^{+0.2}_{-0.2}$     & 0.30$^{+0.09}_{-0.08}$    & 0.21$^{+0.09}_{-0.06}$ \\

    \textsc{gaussian}   &  $E_{\rm Fe}$        & keV            &  6.54$^{+0.01}_{-0.01}$   & 6.53$^{+0.01}_{-0.01}$    & 6.57$^{+0.01}_{-0.01}$    & 6.58$^{+0.01}_{-0.01}$ \\
                                        &  $\sigma_{\rm Fe}$   & keV            &  0.30$^{+0.02}_{-0.02}$   & 0.31$^{+0.02}_{-0.02}$    & 0.28$^{+0.02}_{-0.02}$    & 0.23$^{+0.02}_{-0.02}$ \\
            &  $A_{\rm Fe}$ & $10^{-3}$ ph s$^{-1}$cm$^{-2}$    &  3.4$^{+0.2}_{-0.2}$     & 5.0$^{+0.3}_{-0.3}$        & 2.2$^{+0.1}_{-0.1}$       & 0.96$^{+0.05}_{-0.05}$ \\
        \hline
     $F_{\rm 1-79\,keV}$\tablefootmark{d}       & & $10^{-9}$ erg s$^{-1}$cm$^{-2}$ & 7.49$^{+0.03}_{-0.04}$ & 8.51$^{+0.05}_{-0.04}$ & 4.95$^{+0.01}_{-0.01}$ & 3.082$^{+0.009}_{-0.009}$ \\
     $F_{\rm 4-79\,keV}$\tablefootmark{d}       & & $10^{-9}$ erg s$^{-1}$cm$^{-2}$ & 6.47$^{+0.02}_{-0.02}$ & 7.43$^{+0.03}_{-0.03}$ & 4.183$^{+0.006}_{-0.006}$ & 2.561$^{+0.003}_{-0.003}$ \\
                 $\chi^2$(d.o.f.)  &            &                           & 2085 (1665)                    & 1615 (1564)            & 2963 (2949)                & 3417 (3087)\\ 
    \hline
                 
        \end{tabular}
\end{center}
\tablefoot{
\tablefoottext{a}{Cross-normalisation constant between FPMB and FMPA instruments on board \Nu.}
\tablefoottext{b}{Cross-normalisation constant between \hx/LE and \Nu/FPMA instruments.}
\tablefoottext{c}{Cross-normalisation constant between \hx/ME  and  \Nu/FPMA instruments.}
\tablefoottext{d}{Unabsorbed X-ray flux.}
}
\end{table*}

\subsubsection{Continuum model {\sc comptt}}

The composite model {\sc tbabs$\times$(gau+comptt$\times$gabs)} 
also revealed acceptable fits with $\chi^2$(d.o.f.) values of  2178 (1668), 1756 (1568), 3167 (2952), and 3579 (3089) for four of our data sets. The corresponding fitted spectra are shown in Fig.~\ref{fig:nustar-fit}a$^\prime$--e$^\prime$).
However, we note that the obtained description of the spectra is still not ideal, with some negative residuals around 30~keV clearly seen in all data sets (especially in data set \#3; see Fig.~\ref{fig:nustar-fit}d$^\prime$). This points towards an additional absorption feature, as was already suggested by \cite{Coley2019ATel}. Use of the continuum models such as  {\sc cutoffpl},  {\sc npex}, and {\sc po $\times$ highecut} did not help to get rid of this absorption-like feature (see Fig.~\ref{fig:test-fits}). The fits revealed $\chi^{2}$/d.o.f. of 1731/1527, 1689/1524, and 1933/1526, respectively, for NuObs3, that is, they have not resulted in a significantly improved description of the spectrum. Therefore, it can
be inferred that the residuals are not caused by the use of an inappropriate continuum model. Our best-fit model was then modified by the inclusion of a second Gaussian absorption component {\sc gabs}.

Accordingly, the following spectral analysis was done using model {\sc tbabs$\times$(gau+comptt$\times$gabs$\times$gabs)} which gave better $\chi^{2}$ (d.o.f.) of 2085 (1665), 1615 (1564), 2963 (2949), and 3417 (3087), respectively, for the four data sets in comparison to the one with no absorption model at 30 keV. The corresponding best-fit spectra and residuals are shown in Fig.~\ref{fig:best-fits1} and the best-fit parameters are given in Table~\ref{tab:2}. Unfortunately, the presence of the 30\,keV absorption feature cannot be either confirmed or ruled out based on the \hx\ data, because it appears exactly at the edge of the working energy band of the ME (see, e.g. Fig.~\ref{fig:nustar-fit}). We note that for our four data sets the neutral column density $N_{\rm H}$  varies in the range (0.3--1.4)$\times$ 10$^{22}$ cm$^{-2}$ which is in good agreement with the Galactic mean value of 1.2 $\times$ 10$^{22}$ cm$^{-2}$ in this direction \citep{Willingale2013}. However, it is lower than the value $N_{\rm H}\sim$3.5 $\times$ 10$^{22}$ cm$^{-2}$ previously reported by \citet{Reig2016}, who used the {\sc cutoffpl} model (without the second {\sc gabs}) to fit the spectra obtained during the 2003 outburst.

Finally, in order to estimate the significance of the additional Gaussian absorption component at 30 keV in the \Nu\ observations, we used the {\sc  simftest}\footnote{\url{https://heasarc.gsfc.nasa.gov/xanadu/xspec/manual/node127.html}} script in {\sc xspec}. For this, we simulated $10^4$ spectra and estimated the probability of a false detection. As a result, for all \Nu\ observations fitted by the best-fit model, we obtained the upper limit probability of $10^{-4}$ with the null hypothesis that the absorption component is caused by statistical fluctuations, suggesting the 30-keV feature has a significance of at least $\sim$4$\sigma$.
We also tried to roughly estimate the $F$-test probability using {\sc ftest}\footnote{\url{https://heasarc.gsfc.nasa.gov/xanadu/xspec/manual/node83.html}} task in {\sc xspec} although it may not be a correct method to calculate the significance level of a line-like feature \citep{Protassov2002ApJ}. The $F$-test probability for all four of our  data sets is 4.4 $\times$ $10^{-21}$, 9.6 $\times$ $10^{-27}$, 4.1 $\times$ $10^{-42}$, and 5.7 $\times$ $10^{-32}$, respectively.

In order to investigate the evolution of the {\sc comptt} model on the flux in more detail, we applied it to all 20 \hx/LE+ME spectra, excluding the 30-keV absorption feature from the total model. Interestingly, no absorption feature at 10 keV was required to fit the data in contrast to the {\sc cutoffpl} model. 
However, we let this component remain in the model in order to be able to compare the spectral evolution using both \hx\ and \Nu\ observations. To do this, we fixed the parameters of the 10-keV Gaussian absorption component to the best-fit values obtained from the data set \#3 (NuObs3+P0211006008; see~Table~\ref{tab:2}) representing the average luminosity state of the source.

Figure~\ref{fig:reig-comptt} indicates the evolution of photoelectric absorption ($N_{\rm H}$), plasma temperature ($kT$), input photon temperature ($T_{\rm 0}$), and plasma optical depth ($\tau_{\rm p}$) as a function of flux. As can be seen, the $N_{\rm H}$ value for both \hx\ and \Nu\ observations shows a negative correlation with the X-ray flux varying in the range (0.3--1.5)$\times$ 10$^{22}$ cm$^{-2}$. The $kT$ and $\tau_{\rm p}$ values  do not show clear correlation with the flux. At the same time, the $T_{\rm 0}$ shows a very prominent positive correlation with flux.

Although we used the same composite model to fit the \hx\ and \Nu\ observations, there is some discrepancy in the results obtained with these two observatories, as seen in the figure.
Most probably, the observed differences are caused by the different composition of the data sets from different instruments, as discussed above.

\subsubsection{Two-component continuum model}

Another possible explanation for the observed absorption-like feature around 10 keV is the two-component spectrum. Originally, this idea was proposed for XRP X Persei to explain a similar feature around 20--30 keV \citep{DiSalvo1998,Doroshenko2012-XPer}, which alternatively was also interpreted as a broad CRSF \citep{Coburn2001, Lutovinov2012}. More recently, using observations of two XRPs, A~0535+262 and GX~304--1, in their low-luminosity states, \citet{Tsygankov2019A05,Tsygankov2019-GX} showed that the appearance of the double-hump spectrum is not caused by the cyclotron absorption lines, but may be attributed to the emission of cyclotron photons in the NS atmosphere caused by collisional excitation of electrons to upper Landau levels and consequent Comptonization by hot electron gas \citep{Mushtukov2021}. We emphasise that the two-component spectrum appeared at comparatively low luminosities ($\sim10^{35}$\,erg\,s$^{-1}$) for the sources mentioned above, and so the validity of this scenario is questionable for \source. However, spectra of 1A~0535+262 at higher luminosities ($\sim10^{36}$\,erg\,s$^{-1}$) were found to be well described with the same model \citep{Tsygankov2019A05} even if the presence of both components was not obvious. We note also that a similar shaped spectrum can also be produced by other physical mechanisms, for instance by scattering of the emission from the accretion column of the NS surface \citep{2011A&A...532A..76F}. However, it is important to note that such a complex continuum shape may easily be mistakenly interpreted as a cyclotron absorption feature in the standard XRP spectrum.

In order to verify the applicability of this scenario to \source, we tried to describe the \Nu\ spectra using a model consisting of the two separate components. First, we adopted the same two-component model consisting of two Comptonization models, {\sc comptt,} as previously used to describe the double-hump spectra of XRPs A~0535+262 and GX 304--1 in their low-luminosity states \citep{Tsygankov2019A05,Tsygankov2019-GX}. We then modified it by adding photoelectric absorption ({\sc tbabs}) and an iron line emission component ({\sc gau}).  
For data set \#3, this model resulted in $\chi^2$ = 3306 for 2951 d.o.f., but left residuals around 30 keV. Therefore, we added a Gaussian absorption component ({\sc gabs}) to account for the 30-keV absorption feature which improved the fit significantly to a $\chi^2$ = 2970 for 2948 d.o.f. 

\begin{figure}
\begin{center} 
\includegraphics[width=0.9\columnwidth]{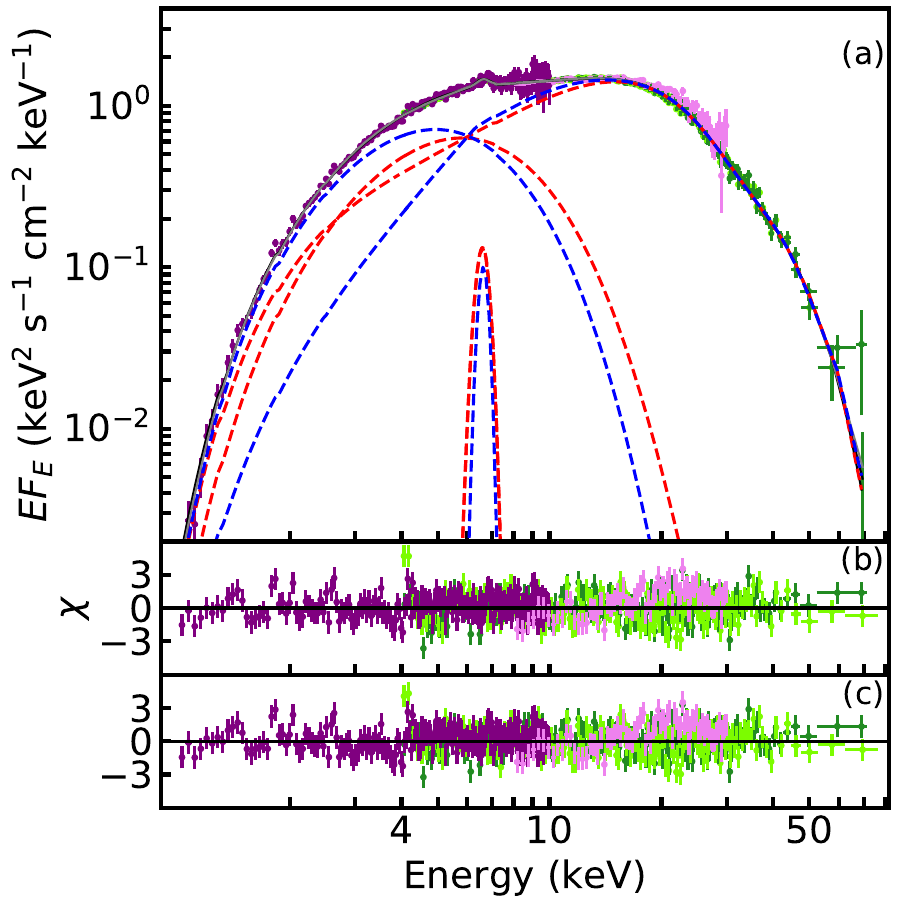}
\end{center}
\caption{\textit{Panel a}: X-ray spectrum of \source\ collected during NuObs3 together with the two-component models {\sc tbabs $\times$ (gau + comptt + comptt $\times$ gabs)} and {\sc tbabs $\times$ (gau + po $\times$ highecut + po $\times$ highecut $\times$ gabs)} where the corresponding model components are shown by the dashed curves in red and blue, respectively. \textit{Panel b}: Residuals for the former composite model ({\sc comptt}). \textit{Panel c}: Residuals for the latter composite model ({\sc po$\times$highecut}). The color code is same as for Fig.~\ref{fig:nustar-fit}.
}

\label{fig:two-compnts} 
\end{figure}

The unfolded spectrum obtained from data set \#3 fitted with a model consisting of two {\sc comptt} components is shown in Fig.~\ref{fig:two-compnts} and the corresponding fit parameters are listed in Table~\ref{tab:3}. As shown in Fig.~\ref{fig:two-compnts}b, the two-component composite model fits the source spectrum without leaving any residuals associated with the absorption feature at 10 keV. Application of this model to the remaining three data sets gives a good fit for NuObs4 ($\chi^2$/d.o.f. = 3901/3087), but leaves some insignificant residuals around 10 keV for the brighter observations NuObs1 ($\chi^2$ (d.o.f.) = 2089 (1663)) and NuObs2 ($\chi^2$/d.o.f. = 1674/1563). The non-ideal quality of the fit can be explained by the fact that the simple {\sc comptt} model is not physically motivated for either component in this kind of spectra. 

To verify this assumption, we adopted the same model that \citet{DiSalvo1998} used to fit the X Persei spectrum consisting of two {\sc (po $\times$ highecut)} components. The modelled spectrum and the corresponding fit parameters for data set \#3 are shown in Fig.~\ref{fig:two-compnts} and Table~\ref{tab:3}. We emphasise that for this composite model,  the 30-keV absorption feature also needs to be taken into consideration. We note that the continuum model {\sc highecut} introduces artificial absorption-line like structures because of the break in the derivative at the cutoff energy. Therefore, in order to get rid of such artificial residuals, we introduced a Gaussian absorption component ({\sc gabs}) to the model \citep[see ][]{Coburn2002} in which the centroid energy is fixed to the position of the cutoff energy determined by {\sc highecut}. The width of the component is also set to 10\% of the centroid energy. This model allowed us to fit all four data sets without leaving any residuals associated with the absorption feature at 10 keV for $\chi^2$(d.o.f.) of 2059 (1664), 1635 (1564), 2949 (2949), and 3492 (3086), respectively. 

\begin{table}
\begin{center}
        \caption{Best-fit parameters from the fit to the NuObs3 spectrum with the two-component models.}
        \label{tab:3}
        \begin{tabular}{lcc} 
        \hline
        \hline
Parameters & \hspace{-1.5cm}Low-energy part & High-energy part              \\
        \hline
\multicolumn{3}{c}{{\sc comptt+comptt}} \\
FMPB/FPMA\tablefootmark{a1}   &     \multicolumn{2}{c}{1.029$\pm$0.001} \\
LE/FPMA\tablefootmark{a2}   &     \multicolumn{2}{c}{0.994$\pm$0.004} \\
ME/FPMA\tablefootmark{a3}   &     \multicolumn{2}{c}{1.005$\pm$0.003} \\
$N_{\rm H}$, $10^{22}$ cm$^{-2}$    &  \multicolumn{2}{c}{$1.5^{+0.1}_{-0.1}$ }         \\
$T_{\rm 0}$, keV                & $0.55^{+0.07}_{-0.07}$             &  $1.416^{+0.054}_{-0.005}$    \\
$kT$, keV                       & $4.63^{+0.05}_{-0.05}$       &  $9.4^{+0.8}_{-0.9}$  \\
$\tau_{\rm p}$                       &  $8.8^{+0.4}_{-0.3}$       &  $<$0.55\tablefootmark{b}   \\
$A_{\rm comptt}$                        &  $0.081^{+0.005}_{-0.005}$ &  $0.025^{+0.001}_{-0.001}$  \\
$E_{\rm abs}$, keV                      & \multicolumn{2}{c}{$32.5^{+0.6}_{-0.6}$} \\
$\sigma_{\rm abs}$, keV             & \multicolumn{2}{c}{$8.8^{+0.7}_{-0.6}$}     \\
$\tau_{\rm abs}$                        & \multicolumn{2}{c}{$0.5^{+0.1}_{-0.1}$} \\
$E_{\rm Fe}$, keV                       &  \multicolumn{2}{c}{6.58$^{+0.01}_{-0.01}$}  \\
$\sigma_{\rm Fe}$, keV              &  \multicolumn{2}{c}{0.26$^{+0.02}_{-0.02}$}   \\
$A_{\rm Fe}$, $10^{-3}$ ph s$^{-1}$cm$^{-2}$    & \multicolumn{2}{c}{2.0$^{+0.1}_{-0.1}$}    \\
$F_{\rm X,1-79\,keV}$, $10^{-9}$ erg s$^{-1}$cm$^{-2}$ & \multicolumn{2}{c}{5.06$\pm$0.05} \\
                 $\chi^2$(d.o.f.)                    & \multicolumn{2}{c}{2970 (2948)}                   \\ \\
\multicolumn{3}{c}{{\sc (po$\times$highecut)+(po$\times$highecut)}} \\ \\
FMPB/FPMA\tablefootmark{a1}   &     \multicolumn{2}{c}{1.029$\pm$0.001} \\
LE/FPMA\tablefootmark{a2}   &     \multicolumn{2}{c}{0.990$\pm$0.004} \\
ME/FPMA\tablefootmark{a3}   &     \multicolumn{2}{c}{1.005$\pm$0.003} \\
$N_{\rm H}$, $10^{22}$ cm$^{-2}$    &  \multicolumn{2}{c}{$1.3^{+0.1}_{-0.1}$}         \\
$\Gamma$                & $-1.8^{+0.3}_{-0.2}$       &  $-0.5^{+0.1}_{-0.3}$    \\
$A_{\rm PL}$, $10^{-2}$  & $2.9^{+1.0}_{-0.6}$       &  $0.7^{+0.2}_{-0.3}$  \\
$E_{\rm cut}$, keV          &  $1.4^{+0.1}_{-0.1}$       &  $6.32^{+0.05}_{-0.06}$   \\
$E_{\rm fold}$, keV         &  $1.24^{+0.23}_{-0.07}$       &  $5.6^{+0.1}_{-0.3}$  \\
$E_{\rm abs}$, keV                      & \multicolumn{2}{c}{$32.3^{+0.7}_{-0.6}$} \\
$\sigma_{\rm abs}$, keV             & \multicolumn{2}{c}{$6.4^{+1.2}_{-0.7}$}     \\
$\tau_{\rm abs}$                            & \multicolumn{2}{c}{$0.3^{+0.2}_{-0.1}$} \\
$E_{\rm Fe}$, keV                       &  \multicolumn{2}{c}{6.62$^{+0.02}_{-0.02}$}  \\
$\sigma_{\rm Fe}$, keV              &  \multicolumn{2}{c}{0.16$^{+0.02}_{-0.02}$}   \\
$A_{\rm Fe}$, $10^{-3}$ ph s$^{-1}$cm$^{-2}$    & \multicolumn{2}{c}{1.2$^{+0.07}_{-0.07}$}    \\
$F_{\rm X,1-79\,keV}$, $10^{-9}$ erg s$^{-1}$cm$^{-2}$ & \multicolumn{2}{c}{5.02$^{+0.02}_{-0.04}$} \\
                 $\chi^2$(d.o.f.)                    & \multicolumn{2}{c}{2949 (2949)}                   \\     
    \hline
        \end{tabular}
\end{center}
\tablefoot{
\tablefoottext{a}{Cross-normalisation constants  FPMB/FPMA \textit{(a1)}, LE/FPMA \textit{(a2)} and ME/FPMA \textit{(a3)}.}
\tablefoottext{b}{3$\sigma$ upper limit.}
}
\end{table}

\begin{table}
    \centering
    \caption{Timing properties of \source. }
    \begin{tabular}{lccc}
    \hline
        \hline
   &  ObsID & $T_{\rm max}$  & $P_{\rm spin}$  \\
   &    & (MJD) & (s)  \\

    \hline

 NuObs1  &  90501305001 & 58531.000014 & 2.762882(2) \\
 NuObs2  &  90502307002 & 58548.9999971 & 2.761549(8)  \\
 \textit{HXMT}  & P021106008 &  58583.2430760 & 2.761858(8)  \\
  NuObs3 & 90502307004 & 58584.0000057 & 2.762115(7)  \\
 NuObs4  & 90501324002 & 58615.0000115 & 2.760750(1)  \\
    \hline
    \end{tabular}
    \label{tab:periods}
\end{table}
 
\subsection{Phase-resolved spectroscopy}
\label{phase-resolved}

\begin{figure}
\begin{center} 
\includegraphics[width=0.9\columnwidth]{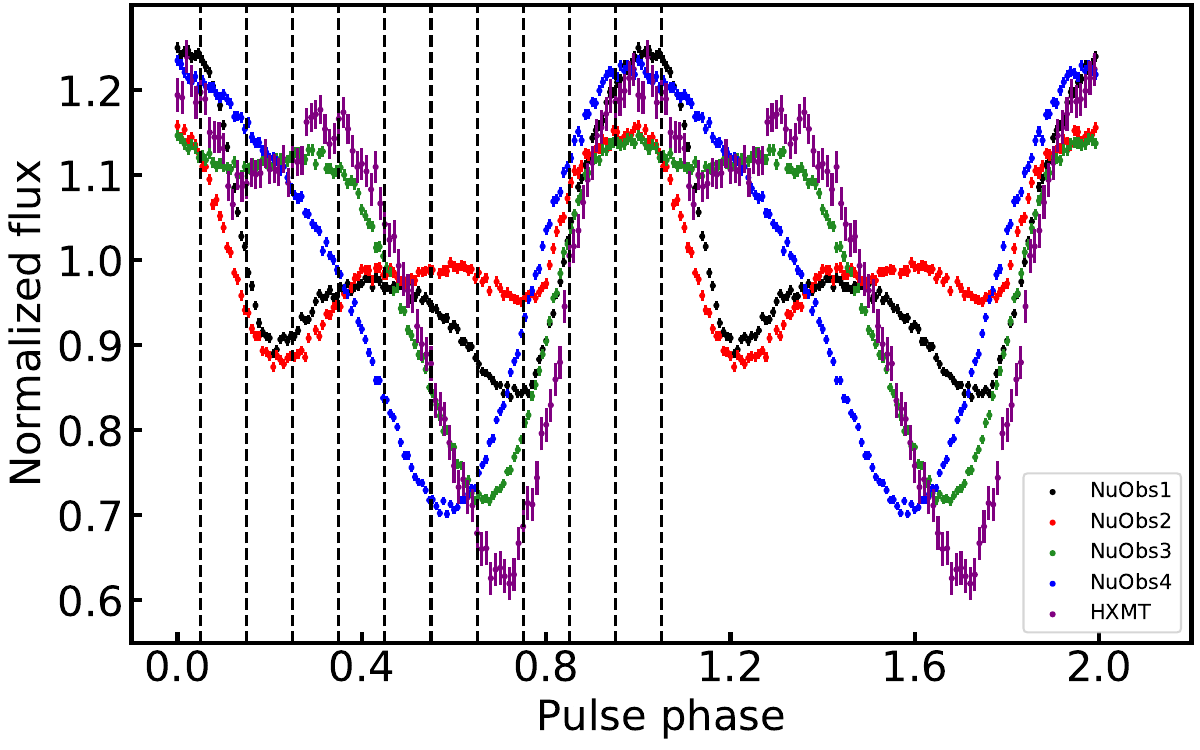}
\end{center}
\caption{Pulse profiles of \source\ from four \Nu\ observations in black (NuObs1), red (NuObs2), green (NuObs3), and blue (NuObs4) in the 3--79 keV energy band. The purple pulse profile was obtained from \hx\ P021106008 in the energy band 1--30 keV. Black dashed lines show the phase segments we used to extract the phase-resolved spectra.
}
\label{fig:pulse-prof} 
\end{figure}

Phase-resolved spectroscopy may serve as a unique source of information about the spatial properties of the emitting region of the NS. Using the barycentrically corrected X-ray light curves, we obtained the spin periods for four \Nu\ observations following the standard epoch folding technique \citep{Leahy1983} using the {\sc efsearch} routines from the {\sc ftool} package (Table~\ref{tab:periods}). The epoch $T_{\rm max}$ of the maximal flux in the pulse profile used as a zero phase for the phase-resolved spectral analysis is presented in the same table for each observation. The spin periods and the corresponding uncertainties were obtained from a simulation of $10^3$ light curves using the count rates and corresponding errors from the original light curves \citep[see e.g.][and references therein]{2013AstL...39..375B}. We then obtained the pulse profiles of all four \Nu\ observations in the full energy band (Fig.~\ref{fig:pulse-prof}). We applied the same procedure to the \hx\ P021106008 observation in order to be able to perform a simultaneous phase-resolved analysis using NuObs3+P021106008. 

A detailed study of the pulse profile evolution of \source\ during the 2019 outburst was performed by \citet{LongJi2020} and \citet{Tuo2020}. We then performed a phase-resolved spectroscopy for \source\ using all four data sets. We find that the dependence of the model parameters on the spin phase is very similar for all the observations and does not depend on the source luminosity. We therefore discuss the results of the phase-resolved analysis mostly for data set \#3 (NuObs3+P021106008) where the 30\,keV feature is most prominent. Taking into the account the available counting statistics, we considered  ten equally spaced phase bins for this analysis as shown in Fig.~\ref{fig:pulse-prof}. Each spectrum was fitted with our best-fit model.
However, we had to fix the width of the 30-keV absorption feature $\sigma_{\rm abs2}$ to its phase-averaged values 5.8 keV as it was not constrained in the fits due to lower count statistics at high energies.

\begin{figure}
\begin{center} 
\includegraphics[width=0.9\columnwidth]{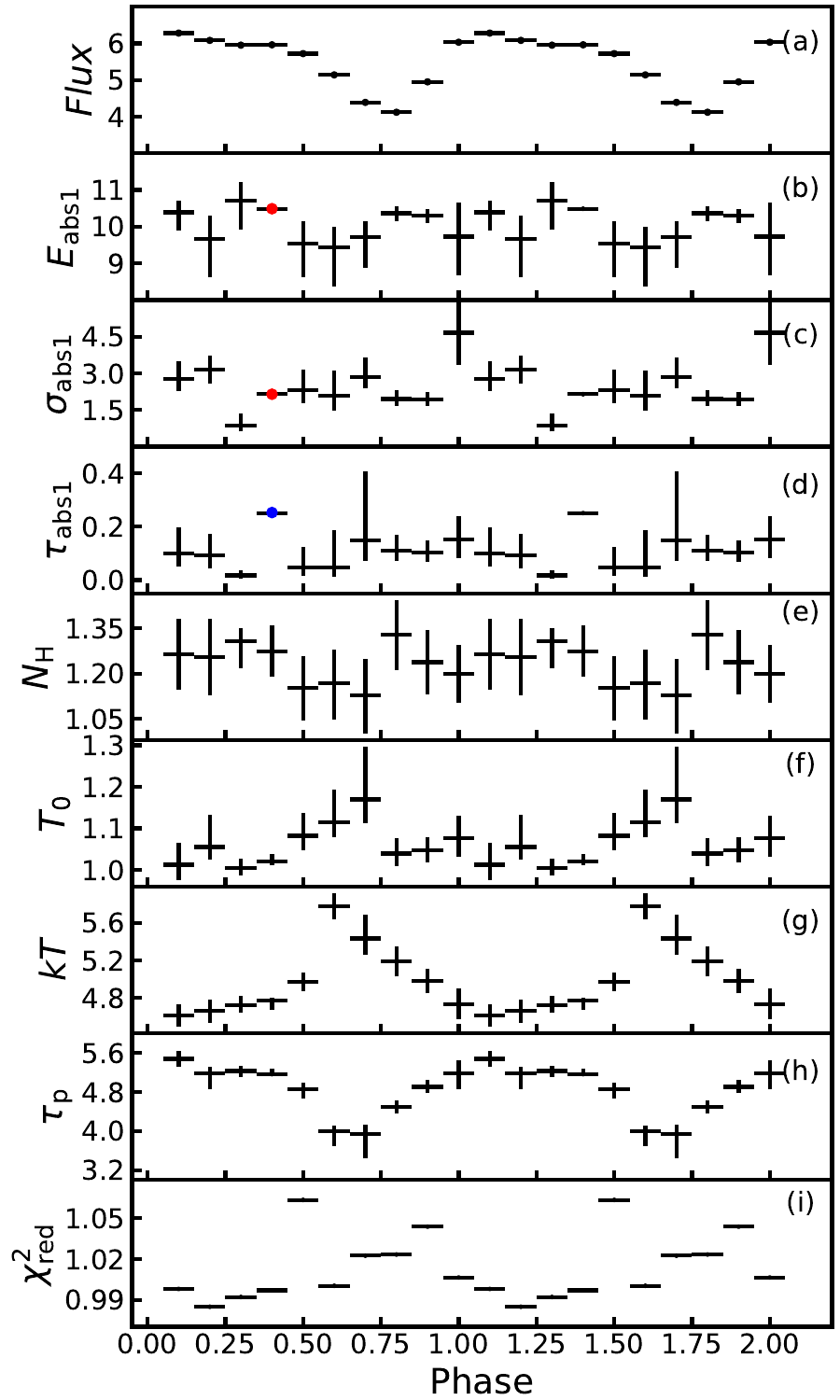}
\end{center}
\caption{Variations of the best-fit parameters of model {\sc tbabs $\times$ (gau+comptt $\times$ gabs $\times$ gabs)} as a function of pulse phase obtained from the NuObs3+P0211006008. Panels \textit{a} to \textit{i} show the observed X-ray flux in the energy band 1--79 keV in units of $10^{-9}$ \flux, first absorption feature energy in keV, width in keV, and its optical depth, neutral hydrogen column density $N_{\rm H}$ in units of $10^{22}$ cm$^{-2}$, input soft photon temperate in keV, plasma temperature in keV, plasma optical depth, and the reduced $\chi^2$, respectively. The values of the red points are fixed at the best-fit values and the blue dot represents the 3$\sigma$ upper limit of the first absorption feature optical depth. The second absorption feature parameters are shown in Fig.~\ref{fig:phase-res_crsf2}.
}
\label{fig:phase-res-1} 
\end{figure}

\begin{figure}
\begin{center} 
\includegraphics[width=0.9\columnwidth]{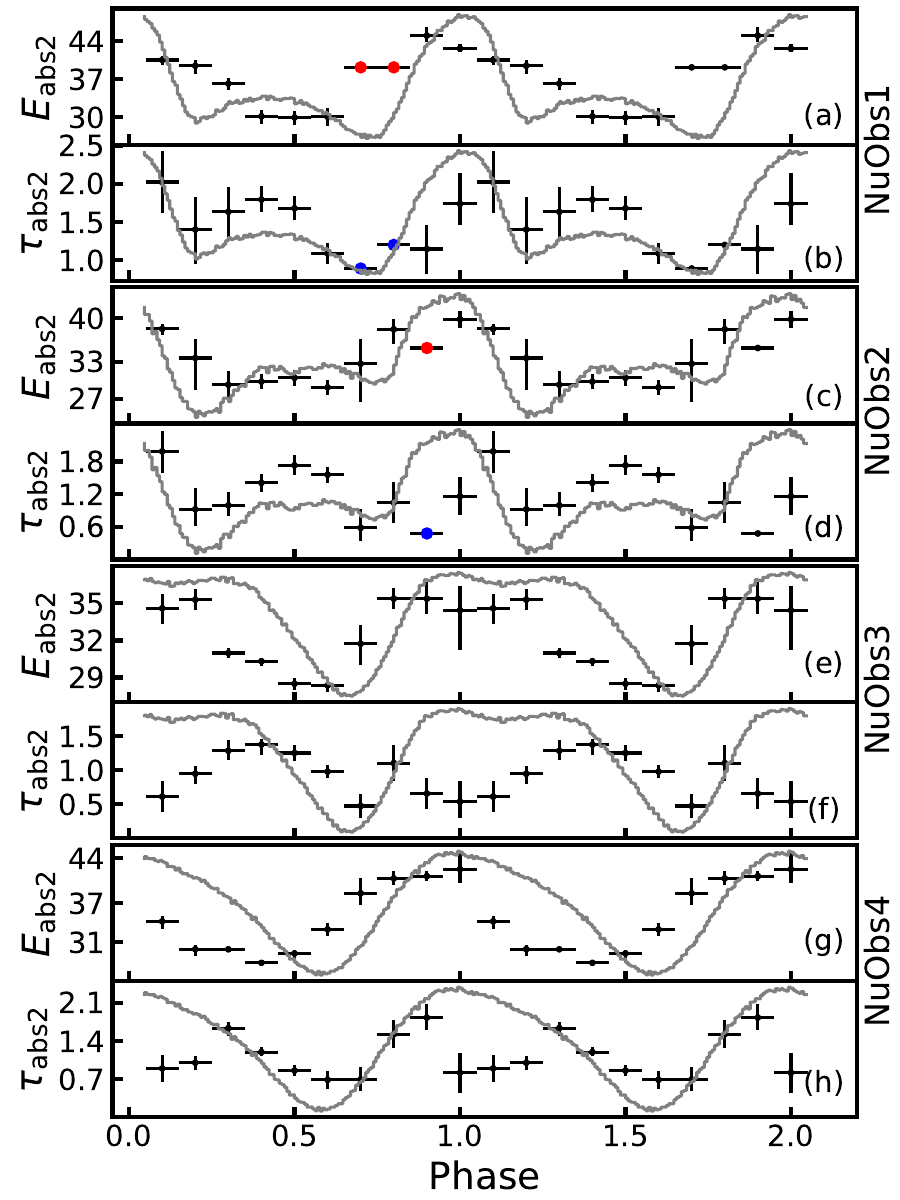}
\end{center}
\caption{Second absorption feature centroid energy (in keV) and its optical depth variations as a function of pulse phase obtained from NuObs1, NuObs2, NuObs3+P0211006008, and NuObs4. The pulse profile of each observation is shown in grey. For each observation, the width of the second absorption feature $\sigma_{\rm abs2}$ was fixed to its phase-averaged value.
}
\label{fig:phase-res_crsf2} 
\end{figure}

The $N_{\rm H}$ remains constant around 1.2 $\times$ 10$^{22}$ cm$^{-2}$ , indicating no dependence on the pulse phase (see Fig.\,\ref{fig:phase-res-1}). The soft photon temperature $T_{0}$ shows insignificant variations between 1 and 1.2 keV. The plasma temperature $kT$ and the optical depth $\tau_{\rm p}$ are clearly dependent on the pulse phases, showing negative and positive correlations with the X-ray flux, respectively. The position $E_{\rm abs1}$ and width $\sigma_{\rm abs1}$ of the first absorption feature remain almost constant around 10 keV and 2-3 keV, respectively. In contrast to that, parameters of the 30-keV absorption feature strongly depend on the pulse phase. In order to illustrate how these variations depend on the source luminosity, we show in Fig.~\ref{fig:phase-res_crsf2} the results of our spectral analysis for all \Nu\ data. As seen from the figure, the position of the 30-keV feature $E_{\rm abs2}$ shows a positive correlation with the X-ray flux. At the same time, the optical depth of the line demonstrates different behaviours at different luminosities. In the first two bright observations, the optical depth is almost correlated with X-ray flux. However, in the latter two observations, it shows a negative correlation with flux in the main maximum and a positive correlation in the main minimum.

\section{Discussion and Conclusions}
\label{discussion}
Using the broadband \Nustar\ data, we find that both the 10 and 30 keV features appear significant when applying conventional continuum models. However, in contrast to previous studies  \citep{Reig2016,2021MNRAS.500.1350B}, we favour the adoption of the Comptonization model {\sc comptt} to describe the broadband X-ray spectrum of the source instead of the {\sc cutoffpl} continuum model as we find this latter to be affected by systematic bias as explained below. 

Particularly, by using {\sc cutoffpl} we were able to confirm that the energy $E_{\rm abs1}$ of the  10
keV feature demonstrates a positive correlation with the X-ray flux as was previously reported for the \textit{RXTE} data \citep[see the Fig. 7 in][]{Reig2016}. This implies that, assuming the 10 keV feature is a CRSF, \source\ remains in a subcritical accretion regime during the outburst with the luminosity below the critical one  \citep{Basko1976,Becker2012,Mushtukov2015-447} which was also discussed by \citet{Reig2016}. Assuming critical luminosity to be around $10^{37}$ \lum, the distance to the source cannot be more than $\sim$3--4 kpc, which is consistent with the \textit{Gaia} measurements \citep{BailerJones2018}. Taking this distance into account, the X-ray luminosities  observed  by \Nustar\  for the source lie in the range (2.8--8.3) $\times$ 10$^{36}$ \lum. Therefore, interpretation of the 10-keV feature as a CRSF in the subcritical pulsar requires a small distance to the source of below $\sim$4~kpc.

However, a more favourable large distance above 12 kpc \citep{Strader2019ATel,Reig2016,Galloway2005,Tuo2020} results in X-ray luminosities in the range (4.5--13.3) $\times$ 10$^{37}$ \lum\ for the four \Nu\ observations, which is above the critical luminosity for a XRP with a typical magnetic field. This is at odds with the idea that the 10-keV feature has a magnetic origin because, in the case of bright XRPs, a negative CRSF energy--luminosity correlation is expected \citep{Mihara1998, Tsygankov2006, Becker2012}. Moreover, \cite{Mushtukov2015-447} demonstrated that for a XRP with magnetic field around $10^{12}$~G, the critical luminosity may be as low as a few $\times$ 10$^{36}$ \lum. In this case, according to the \Nu\ observations, \source\ should be in the supercritical accretion regime during the outburst even at a distance of 3 kpc. Therefore, the observed properties of the 10-keV feature (i.e. positive correlation of $E_{\rm abs1}$ with X-ray luminosity) is hard to explain in terms of the cyclotron line for any available estimates of the distance. It is worth noting that the \Nu\ spectra may show an instrumental feature around 10 keV as seen in some cases. For instance, in Be-transient XRP KS 1947+300, the 12.5 keV absorption feature was long debated until recently when \citet{Doroshenko-R2020} drew attention to the instrumental effects causing the appearance of  the feature in the \Nu\ spectra. The instrumental systematic effect in \Nu\ observations results from the tungsten L-edge of the telescope optics coupled with small uncertainties in the energy scale offset \citep{Madsen-NuSTAR2015}. However, the existence of the absorption like feature around 10 keV is confirmed by the \textit{RXTE}/PCA \citep{Reig2016}, ruling out an instrumental origin.

Moreover, we noticed that the combination of the continuum model {\sc cutoffpl} with a Gaussian absorption component {\sc gabs} tends to modify the continuum itself rather than  producing a narrow absorption feature at 10 keV. As a result, the width of the feature at low flux becomes comparable to its centroid energy (see Figs. \ref{fig:reig} and \ref{fig:gabs-anomoly}). This behaviour was not discussed previously in the literature and casts serious doubt on the CRSF nature of the feature.

Taking into account all the problems mentioned above, we applied {\sc comptt} to the continuum, which resulted in a significantly different description of the spectrum. Particularly, we did not detect an absorption feature at 10 keV in any of the 20 \hx\ observations. This is in line with the results of \cite{Reig2016} who also admitted that the Comptonization models cause the 10-keV absorption feature to appear insignificant in the \textit{RXTE}/PCA data.
However, the feature was required in all \Nu\ observations. Unfortunately, the Comptonization model also showed the same problem as {\sc cutoffpl} by revealing a large width $\sigma_{\rm abs1}$ = 4.4 keV for the centroid energy $E_{\rm abs1}$ = 9 keV in the NuObs4 which was obtained when the source showed low luminosity. Interestingly, the centroid energy $E_{\rm abs1}$ of the  10 keV feature remains almost constant at around 10.6 keV during the outburst as seen in the first three \Nu\ observations. This  clearly demonstrates that some properties of the 10 keV feature depend on the continuum model.

We demonstrate that the position of the 10 keV feature is stable along the pulse phase.
On the other hand, phase-resolved spectroscopy reveals a strong dependence of the 30 keV feature on the pulse phases in all luminosity states of the source, which is typical behaviour for CRSFs. It is worth mentioning here that the centroid energies of the two absorption features do not appear to be harmonically related. Although not a typical behaviour for CRSFs, this can be explained by assuming that both features are associated with fundamental cyclotron lines originating at different altitudes above the NS. Particularly, considering the strong dependence of the magnetic field strength on the height of the emission region ($B\propto h^{-3}$), the 10 keV absorption line may originate at $\sim$4.5 km (for a NS radius of $R_{*}$ = 10 km) above the NS surface. This scenario may also be combined with the so-called offset magnetic dipole
resulting in the presence of two independent fundamental cyclotron lines \citep[see e.g.][]{Iyer2015}.

Finally, we suggest another possible explanation for the 10 keV absorption feature. We demonstrate that the two-component continuum provides an acceptable description of the broadband spectra of \source\ with no need to introduce an absorption 10 keV feature into the model, despite the two-component {\sc comptt} model leaving some residuals around 10 keV in three out of four data sets. The purpose of using {\sc comptt} in this context as a phenomenological model was to be able to compare the results with the previous observational studies. However, physically motivated theoretical models for the emission from highly magnetised NS atmospheres at low mass-accretion rates have appeared recently and demonstrate very good consistency with the observational data on other low-luminosity XRPs \citep{Mushtukov2021,Sokolova-Lapa2021}.

It should be noted that transformation of a typical cutoff power-law-shaped spectrum observed in XRPs at high luminosities to a double-component spectrum happens at low luminosities of $\sim$10$^{34-36}$ \lum\ \citep{Doroshenko2012-XPer,Tsygankov2019A05,Tsygankov2019-GX,2021ApJ...912...17L}. Therefore, this explanation may also require a small distance to the source. At the same time, evidence for a two-component continuum starts to appear in some bright XRPs as well \citep{Doroshenko-R2020}. We emphasise that even in the case of the two-component continuum model, the 30-keV feature still needs to be taken into account.

It is worth noting that a magnetic origin of the 30-keV feature in the source spectrum was supported by the investigation of the pulse profile shape as a function of energy performed by \cite{2021MNRAS.500.1350B}. The authors discovered an abrupt change in the pulse profiles around this energy that is a typical behaviour around CRSF energy \citep[see e.g.][]{Tsygankov2006}. At the same time, no significant deviation from the general trend around 10 keV was found. This fact also supports speculation that the 10 keV feature may have a non-magnetic origin.

\begin{acknowledgements}
This work was supported by the grant 14.W03.31.0021 of the Ministry of Science and Higher Education of the Russian Federation.
We also acknowledge the support from the Finnish Cultural Foundation project 00200764 (AN), the Academy of Finland travel grants 317552, 322779, 324550, 331951, and 333112 (AN, SST, JP), the German Academic Exchange Service (DAAD) travel grants 57405000 and 57525212 (LJ, VD), the National Key R\&D Program of China (2016YFA0400800), and the National Natural Science Foundation of China grants U1838201, U1838202, U1938103 and 11733009 (LJ, YL, SZ, SZ, LJ). 
This work made use of data from the \textit{Insight}--HXMT mission, a project funded by China National Space Administration (CNSA) and the Chinese Academy of Sciences (CAS).
\end{acknowledgements}

\bibliographystyle{aa}
\bibliography{library}

\begin{thebibliography}{65}
\expandafter\ifx\csname natexlab\endcsname\relax\def\natexlab#1{#1}\fi

\bibitem[{{Alpar} \& {Shaham}(1985)}]{Alpar-Shaham1985}
{Alpar}, M.~A. \& {Shaham}, J. 1985, \nat, 316, 239

\bibitem[{{Anders} \& {Grevesse}(1989)}]{An-Gre1989}
{Anders}, E. \& {Grevesse}, N. 1989, \gca, 53, 197

\bibitem[{{Bailer-Jones} {et~al.}(2018){Bailer-Jones}, {Rybizki}, {Fouesneau},
  {Mantelet}, \& {Andrae}}]{BailerJones2018}
{Bailer-Jones}, C.~A.~L., {Rybizki}, J., {Fouesneau}, M., {Mantelet}, G., \&
  {Andrae}, R. 2018, \aj, 156, 58

\bibitem[{{Basko} \& {Sunyaev}(1976)}]{Basko1976}
{Basko}, M.~M. \& {Sunyaev}, R.~A. 1976, \mnras, 175, 395

\bibitem[{{Becker} {et~al.}(2012){Becker}, {Klochkov}, {Sch{\"o}nherr},
  {Nishimura}, {Ferrigno}, {Caballero}, {Kretschmar}, {Wolff}, {Wilms}, \&
  {Staubert}}]{Becker2012}
{Becker}, P.~A., {Klochkov}, D., {Sch{\"o}nherr}, G., {et~al.} 2012, \aap, 544,
  A123

\bibitem[{{Becker} \& {Wolff}(2007)}]{2007ApJ...654..435B}
{Becker}, P.~A. \& {Wolff}, M.~T. 2007, \apj, 654, 435

\bibitem[{{Beri} {et~al.}(2021){Beri}, {Girdhar}, {Iyer}, \&
  {Maitra}}]{2021MNRAS.500.1350B}
{Beri}, A., {Girdhar}, T., {Iyer}, N.~K., \& {Maitra}, C. 2021, \mnras, 500,
  1350

\bibitem[{{Boldin} {et~al.}(2013){Boldin}, {Tsygankov}, \&
  {Lutovinov}}]{2013AstL...39..375B}
{Boldin}, P.~A., {Tsygankov}, S.~S., \& {Lutovinov}, A.~A. 2013, Astronomy
  Letters, 39, 375

\bibitem[{{Cao} {et~al.}(2020){Cao}, {Jiang}, {Meng}, {Zhang}, {Luo}, {Yang},
  {Zhang}, {Gu}, {Sun}, {Liu}, {Yang}, {Li}, {Tan}, {Liu}, {Du}, {Lu}, {Xu},
  {Guan}, {Zhang}, {Wang}, {Li}, {Zhang}, {Wen}, {Qu}, {Song}, {Li}, {Ge},
  {Zhou}, {Xiong}, {Zhang}, {Zhang}, {Cheng}, {Zhang}, {Li}, {Liang}, {Gao},
  {Yang}, {Liu}, {Liu}, {Yang}, \& {Zhang}}]{Cao2020}
{Cao}, X., {Jiang}, W., {Meng}, B., {et~al.} 2020, Science China Physics,
  Mechanics, and Astronomy, 63, 249504

\bibitem[{{Chen} {et~al.}(2008){Chen}, {Qu}, {Zhang}, {Zhang}, \&
  {Zhang}}]{Chen2008}
{Chen}, W., {Qu}, J.-l., {Zhang}, S., {Zhang}, F., \& {Zhang}, G.-b. 2008,
  \caa, 32, 241

\bibitem[{{Chen} {et~al.}(2020){Chen}, {Cui}, {Li}, {Wang}, {Xu}, {Lu}, {Wang},
  {Chen}, {Han}, {Hu}, {Zhang}, {Huo}, {Yang}, {Li}, {Lu}, {Zhang}, {Li},
  {Zhang}, {Xiong}, {Zhang}, {Xue}, {Zhao}, {Zhu}, {Zhu}, {Liu}, {Yang}, \&
  {Zhang}}]{Chen2020}
{Chen}, Y., {Cui}, W., {Li}, W., {et~al.} 2020, Science China Physics,
  Mechanics, and Astronomy, 63, 249505

\bibitem[{{Coburn} {et~al.}(2001){Coburn}, {Heindl}, {Gruber}, {Rothschild},
  {Staubert}, {Wilms}, \& {Kreykenbohm}}]{Coburn2001}
{Coburn}, W., {Heindl}, W.~A., {Gruber}, D.~E., {et~al.} 2001, \apj, 552, 738

\bibitem[{{Coburn} {et~al.}(2002){Coburn}, {Heindl}, {Rothschild}, {Gruber},
  {Kreykenbohm}, {Wilms}, {Kretschmar}, \& {Staubert}}]{Coburn2002}
{Coburn}, W., {Heindl}, W.~A., {Rothschild}, R.~E., {et~al.} 2002, \apj, 580,
  394

\bibitem[{{Coley} {et~al.}(2019){Coley}, {Fuerst}, {Hemphill}, {Kretschmar},
  {Pottschmidt}, {Jaisawal}, {Malacaria}, {Vasilopoulos}, {Wilms}, \&
  {Wolff}}]{Coley2019ATel}
{Coley}, J.~B., {Fuerst}, F., {Hemphill}, P., {et~al.} 2019, The Astronomer's
  Telegram, 12684, 1

\bibitem[{{Di Salvo} {et~al.}(1998){Di Salvo}, {Burderi}, {Robba}, \&
  {Guainazzi}}]{DiSalvo1998}
{Di Salvo}, T., {Burderi}, L., {Robba}, N.~R., \& {Guainazzi}, M. 1998, \apj,
  509, 897

\bibitem[{{Doroshenko} {et~al.}(2020){Doroshenko}, {Piraino}, {Doroshenko}, \&
  {Santangelo}}]{Doroshenko-R2020}
{Doroshenko}, R., {Piraino}, S., {Doroshenko}, V., \& {Santangelo}, A. 2020,
  \mnras, 493, 3442

\bibitem[{{Doroshenko} {et~al.}(2012){Doroshenko}, {Santangelo}, {Kreykenbohm},
  \& {Doroshenko}}]{Doroshenko2012-XPer}
{Doroshenko}, V., {Santangelo}, A., {Kreykenbohm}, I., \& {Doroshenko}, R.
  2012, \aap, 540, L1

\bibitem[{{Farinelli} {et~al.}(2012){Farinelli}, {Ceccobello}, {Romano}, \&
  {Titarchuk}}]{2012A&A...538A..67F}
{Farinelli}, R., {Ceccobello}, C., {Romano}, P., \& {Titarchuk}, L. 2012, \aap,
  538, A67

\bibitem[{{Ferrigno} {et~al.}(2011){Ferrigno}, {Falanga}, {Bozzo}, {Becker},
  {Klochkov}, \& {Santangelo}}]{2011A&A...532A..76F}
{Ferrigno}, C., {Falanga}, M., {Bozzo}, E., {et~al.} 2011, \aap, 532, A76

\bibitem[{{Filippova} {et~al.}(2005){Filippova}, {Tsygankov}, {Lutovinov}, \&
  {Sunyaev}}]{Filippova2005}
{Filippova}, E.~V., {Tsygankov}, S.~S., {Lutovinov}, A.~A., \& {Sunyaev}, R.~A.
  2005, Astronomy Letters, 31, 729

\bibitem[{{Forman} {et~al.}(1976){Forman}, {Jones}, \&
  {Tananbaum}}]{1976ApJ...206L..29F}
{Forman}, W., {Jones}, C., \& {Tananbaum}, H. 1976, \apjl, 206, L29

\bibitem[{{Galloway} {et~al.}(2003){Galloway}, {Remillard}, {Morgan}, \&
  {Swank}}]{Galloway2003b}
{Galloway}, D., {Remillard}, R., {Morgan}, E., \& {Swank}, J. 2003, \iaucirc,
  8070, 2

\bibitem[{{Galloway} {et~al.}(2005){Galloway}, {Wang}, \&
  {Morgan}}]{Galloway2005}
{Galloway}, D.~K., {Wang}, Z., \& {Morgan}, E.~H. 2005, \apj, 635, 1217

\bibitem[{{Gornostaev}(2021)}]{2021MNRAS.501..564G}
{Gornostaev}, M.~I. 2021, \mnras, 501, 564

\bibitem[{{Green} {et~al.}(2018){Green}, {Schlafly}, {Finkbeiner}, {Rix},
  {Martin}, {Burgett}, {Draper}, {Flewelling}, {Hodapp}, {Kaiser}, {Kudritzki},
  {Magnier}, {Metcalfe}, {Tonry}, {Wainscoat}, \&
  {Waters}}]{2018MNRAS.478..651G}
{Green}, G.~M., {Schlafly}, E.~F., {Finkbeiner}, D., {et~al.} 2018, \mnras,
  478, 651

\bibitem[{{Guo} {et~al.}(2020){Guo}, {Liao}, {Zhang}, {Zhang}, {Tan}, {Song},
  {Lu}, {Cao}, {Chang}, {Chen}, {Du}, {Ge}, {Gu}, {Jiang}, {Jin}, {Li}, {Li},
  {Li}, {Liu}, {Liu}, {Lu}, {Luo}, {Meng}, {Sun}, {Yang}, {Yang}, {You},
  {Zhang}, {Zhao}, \& {Zhang}}]{Guo2020}
{Guo}, C.-C., {Liao}, J.-Y., {Zhang}, S., {et~al.} 2020, Journal of High Energy
  Astrophysics, 27, 44

\bibitem[{{Harrison} {et~al.}(2013){Harrison}, {Craig}, {Christensen},
  {Hailey}, {Zhang}, {Boggs}, {Stern}, {Cook}, {Forster}, {Giommi},
  {Grefenstette}, {Kim}, {Kitaguchi}, {Koglin}, {Madsen}, {Mao}, {Miyasaka},
  {Mori}, {Perri}, {Pivovaroff}, {Puccetti}, {Rana}, {Westergaard}, {Willis},
  {Zoglauer}, {An}, {Bachetti}, {Barri{\`e}re}, {Bellm}, {Bhalerao},
  {Brejnholt}, {Fuerst}, {Liebe}, {Markwardt}, {Nynka}, {Vogel}, {Walton},
  {Wik}, {Alexander}, {Cominsky}, {Hornschemeier}, {Hornstrup}, {Kaspi},
  {Madejski}, {Matt}, {Molendi}, {Smith}, {Tomsick}, {Ajello}, {Ballantyne},
  {Balokovi{\'c}}, {Barret}, {Bauer}, {Blandford}, {Brandt}, {Brenneman},
  {Chiang}, {Chakrabarty}, {Chenevez}, {Comastri}, {Dufour}, {Elvis}, {Fabian},
  {Farrah}, {Fryer}, {Gotthelf}, {Grindlay}, {Helfand}, {Krivonos}, {Meier},
  {Miller}, {Natalucci}, {Ogle}, {Ofek}, {Ptak}, {Reynolds}, {Rigby},
  {Tagliaferri}, {Thorsett}, {Treister}, \& {Urry}}]{Harrison2013}
{Harrison}, F.~A., {Craig}, W.~W., {Christensen}, F.~E., {et~al.} 2013, \apj,
  770, 103

\bibitem[{{Hemphill} {et~al.}(2019){Hemphill}, {Coley}, {Fuerst}, {Kretschmar},
  {Kuehnel}, {Malacaria}, \& {Pottschmidt}}]{Hemphill2019ATel}
{Hemphill}, P., {Coley}, J., {Fuerst}, F., {et~al.} 2019, The Astronomer's
  Telegram, 12556, 1

\bibitem[{{Iyer} {et~al.}(2015){Iyer}, {Mukherjee}, {Dewangan}, {Bhattacharya},
  \& {Seetha}}]{Iyer2015}
{Iyer}, N., {Mukherjee}, D., {Dewangan}, G.~C., {Bhattacharya}, D., \&
  {Seetha}, S. 2015, \mnras, 454, 741

\bibitem[{{James} {et~al.}(2011){James}, {Paul}, {Devasia}, \&
  {Indulekha}}]{James2011}
{James}, M., {Paul}, B., {Devasia}, J., \& {Indulekha}, K. 2011, \mnras, 410,
  1489

\bibitem[{{Ji} {et~al.}(2020){Ji}, {Ducci}, {Santangelo}, {Zhang},
  {Suleimanov}, {Tsygankov}, {Doroshenko}, {Nabizadeh}, {Zhang}, {Ge}, {Tao},
  {Bu}, {Qu}, {Lu}, {Chen}, {Song}, {Li}, {Xu}, {Cao}, {Chen}, {Liu}, {Cai},
  {Chang}, {Chen}, {Chen}, {Chen}, {Chen}, {Cui}, {Cui}, {Deng}, {Dong}, {Du},
  {Fu}, {Gao}, {Gao}, {Gao}, {Gu}, {Guan}, {Guo}, {Han}, {Huang}, {Huo}, {Jia},
  {Jiang}, {Jiang}, {Jin}, {Jin}, {Kong}, {Li}, {Li}, {Li}, {Li}, {Li}, {Li},
  {Li}, {Li}, {Li}, {Li}, {Liang}, {Liao}, {Liu}, {Liu}, {Liu}, {Liu}, {Liu},
  {Liu}, {Lu}, {Lu}, {Luo}, {Luo}, {Ma}, {Meng}, {Nang}, {Nie}, {Ou}, {Sai},
  {Shang}, {Song}, {Sun}, {Tan}, {Tuo}, {Wang}, {Wang}, {Wang}, {Wang}, {Wang},
  {Wang}, {Wen}, {Wu}, {Wu}, {Wu}, {Xiao}, {Xiao}, {Xiong}, {Xu}, {Yang},
  {Yang}, {Yang}, {Yang}, {Yi}, {Yin}, {You}, {Zhang}, {Zhang}, {Zhang},
  {Zhang}, {Zhang}, {Zhang}, {Zhang}, {Zhang}, {Zhang}, {Zhang}, {Zhang},
  {Zhang}, {Zhang}, {Zhang}, {Zhang}, {Zhang}, {Zhang}, {Zhang}, {Zhao},
  {Zhao}, {Zheng}, {Zhou}, {Zhou}, {Zhu}, {Zhu}, \& {Zhuang}}]{LongJi2020}
{Ji}, L., {Ducci}, L., {Santangelo}, A., {et~al.} 2020, \mnras, 493, 5680

\bibitem[{{Leahy} {et~al.}(1983){Leahy}, {Darbro}, {Elsner}, {Weisskopf},
  {Sutherland}, {Kahn}, \& {Grindlay}}]{Leahy1983}
{Leahy}, D.~A., {Darbro}, W., {Elsner}, R.~F., {et~al.} 1983, \apj, 266, 160

\bibitem[{{Lei} {et~al.}(2009){Lei}, {Chen}, {Qu}, {Song}, {Zhang}, {Lu},
  {Zhang}, \& {Li}}]{Lei2009}
{Lei}, Y.-J., {Chen}, W., {Qu}, J.-L., {et~al.} 2009, \apj, 707, 1016

\bibitem[{{Li}(2007)}]{LI2007131}
{Li}, T.-P. 2007, Nuclear Physics B Proceedings Supplements, 166, 131

\bibitem[{{Li} {et~al.}(2020){Li}, {Li}, {Tan}, {Yang}, {Ge}, {Zhang}, {Tuo},
  {Wu}, {Liao}, {Zhang}, {Song}, {Zhang}, {Qu}, {Zhang}, {Lu}, {Xu}, {Liu},
  {Cao}, {Chen}, {Nie}, {Zhao}, \& {Li}}]{Li2020}
{Li}, X., {Li}, X., {Tan}, Y., {et~al.} 2020, Journal of High Energy
  Astrophysics, 27, 64

\bibitem[{{Liao} {et~al.}(2020{\natexlab{a}}){Liao}, {Zhang}, {Chen}, {Zhang},
  {Jin}, {Chang}, {Chen}, {Ge}, {Guo}, {Li}, {Li}, {Lu}, {Lu}, {Nie}, {Song},
  {Yang}, {You}, {Zhao}, \& {Zhang}}]{Liao2020b}
{Liao}, J.-Y., {Zhang}, S., {Chen}, Y., {et~al.} 2020{\natexlab{a}}, Journal of
  High Energy Astrophysics, 27, 24

\bibitem[{{Liao} {et~al.}(2020{\natexlab{b}}){Liao}, {Zhang}, {Lu}, {Zhang},
  {Li}, {Chang}, {Chen}, {Ge}, {Guo}, {Huang}, {Jin}, {Li}, {Li}, {Li}, {Liu},
  {Lu}, {Nie}, {Song}, {Wang}, {You}, {Zhang}, {Zhao}, \& {Zhang}}]{Liao2020}
{Liao}, J.-Y., {Zhang}, S., {Lu}, X.-F., {et~al.} 2020{\natexlab{b}}, Journal
  of High Energy Astrophysics, 27, 14

\bibitem[{{Liu} {et~al.}(2020){Liu}, {Zhang}, {Li}, {Lu}, {Chang}, {Li},
  {Zhang}, {Jin}, {Yu}, {Zhang}, {Fu}, {Chen}, {Ji}, {Xu}, {Deng}, {Shang},
  {Liu}, {Lu}, {Zhang}, {Dong}, {Li}, {Wu}, {Li}, {Wang}, {Wu}, {Zhang},
  {Zhang}, {Xiong}, {Liu}, {Zhang}, {Liu}, {Yang}, \& {Zhang}}]{Liu2020}
{Liu}, C., {Zhang}, Y., {Li}, X., {et~al.} 2020, Science China Physics,
  Mechanics, and Astronomy, 63, 249503

\bibitem[{{Lutovinov} {et~al.}(2012){Lutovinov}, {Tsygankov}, \&
  {Chernyakova}}]{Lutovinov2012}
{Lutovinov}, A., {Tsygankov}, S., \& {Chernyakova}, M. 2012, \mnras, 423, 1978

\bibitem[{{Lutovinov} {et~al.}(2021){Lutovinov}, {Tsygankov}, {Molkov},
  {Doroshenko}, {Mushtukov}, {Arefiev}, {Lapshov}, {Tkachenko}, \&
  {Pavlinsky}}]{2021ApJ...912...17L}
{Lutovinov}, A., {Tsygankov}, S., {Molkov}, S., {et~al.} 2021, \apj, 912, 17

\bibitem[{{Madsen} {et~al.}(2015){Madsen}, {Harrison}, {Markwardt}, {An},
  {Grefenstette}, {Bachetti}, {Miyasaka}, {Kitaguchi}, {Bhalerao}, {Boggs},
  {Christensen}, {Craig}, {Forster}, {Fuerst}, {Hailey}, {Perri}, {Puccetti},
  {Rana}, {Stern}, {Walton}, {J{\o}rgen Westergaard}, \&
  {Zhang}}]{Madsen-NuSTAR2015}
{Madsen}, K.~K., {Harrison}, F.~A., {Markwardt}, C.~B., {et~al.} 2015, \apjs,
  220, 8

\bibitem[{{McCollum} \& {Laine}(2019)}]{McCollum2019ATel}
{McCollum}, B. \& {Laine}, S. 2019, The Astronomer's Telegram, 12560, 1

\bibitem[{{Mereminskiy} {et~al.}(2019){Mereminskiy}, {Lutovinov}, {Tsygankov},
  {Semena}, \& {Shtykovskiy}}]{2019ATel12514....1M}
{Mereminskiy}, I.~A., {Lutovinov}, A.~A., {Tsygankov}, S.~S., {Semena}, A.~N.,
  \& {Shtykovskiy}, A.~E. 2019, The Astronomer's Telegram, 12514, 1

\bibitem[{{Mihara} {et~al.}(1998){Mihara}, {Makishima}, \&
  {Nagase}}]{Mihara1998}
{Mihara}, T., {Makishima}, K., \& {Nagase}, F. 1998, Advances in Space
  Research, 22, 987

\bibitem[{{Miller} {et~al.}(1998){Miller}, {Lamb}, \& {Psaltis}}]{Miller1998}
{Miller}, M.~C., {Lamb}, F.~K., \& {Psaltis}, D. 1998, \apj, 508, 791

\bibitem[{{Mushtukov} {et~al.}(2021){Mushtukov}, {Suleimanov}, {Tsygankov}, \&
  {Portegies Zwart}}]{Mushtukov2021}
{Mushtukov}, A.~A., {Suleimanov}, V.~F., {Tsygankov}, S.~S., \& {Portegies
  Zwart}, S. 2021, \mnras, 503, 5193

\bibitem[{{Mushtukov} {et~al.}(2015){Mushtukov}, {Suleimanov}, {Tsygankov}, \&
  {Poutanen}}]{Mushtukov2015-447}
{Mushtukov}, A.~A., {Suleimanov}, V.~F., {Tsygankov}, S.~S., \& {Poutanen}, J.
  2015, \mnras, 447, 1847

\bibitem[{{Priedhorsky} \& {Terrell}(1984)}]{1984ApJ...280..661P}
{Priedhorsky}, W.~C. \& {Terrell}, J. 1984, \apj, 280, 661

\bibitem[{{Protassov} {et~al.}(2002){Protassov}, {van Dyk}, {Connors},
  {Kashyap}, \& {Siemiginowska}}]{Protassov2002ApJ}
{Protassov}, R., {van Dyk}, D.~A., {Connors}, A., {Kashyap}, V.~L., \&
  {Siemiginowska}, A. 2002, \apj, 571, 545

\bibitem[{{Reig} \& {Milonaki}(2016)}]{Reig2016}
{Reig}, P. \& {Milonaki}, F. 2016, \aap, 594, A45

\bibitem[{{Sokolova-Lapa} {et~al.}(2021){Sokolova-Lapa}, {Gornostaev}, {Wilms},
  {Ballhausen}, {Falkner}, {Postnov}, {Thalhammer}, {F{\"u}rst}, {Garc{\'\i}a},
  {Shakura}, {Becker}, {Wolff}, {Pottschmidt}, {H{\"a}rer}, \&
  {Malacaria}}]{Sokolova-Lapa2021}
{Sokolova-Lapa}, E., {Gornostaev}, M., {Wilms}, J., {et~al.} 2021, arXiv
  e-prints, arXiv:2104.06802

\bibitem[{{Staubert} {et~al.}(2019){Staubert}, {Tr{\"u}mper}, {Kendziorra},
  {Klochkov}, {Postnov}, {Kretschmar}, {Pottschmidt}, {Haberl}, {Rothschild},
  {Santangelo}, {Wilms}, {Kreykenbohm}, \& {F{\"u}rst}}]{Staubert2019}
{Staubert}, R., {Tr{\"u}mper}, J., {Kendziorra}, E., {et~al.} 2019, \aap, 622,
  A61

\bibitem[{{Strader} {et~al.}(2019){Strader}, {Chomiuk}, {Swihart}, \&
  {Aydi}}]{Strader2019ATel}
{Strader}, J., {Chomiuk}, L., {Swihart}, S., \& {Aydi}, E. 2019, The
  Astronomer's Telegram, 12554, 1

\bibitem[{{Tanaka}(1986)}]{Tanaka1986}
{Tanaka}, Y. 1986, in Lecture Notes in Physics, Vol. 255, Radiation
  Hydrodynamics in Stars and Compact Objects, ed. D.~{Mihalas} \& K.-H.~A.
  {Winkler} (Berlin: Springer-Verlag), 198

\bibitem[{{Titarchuk}(1994)}]{Titarchuk1994}
{Titarchuk}, L. 1994, \apj, 434, 570

\bibitem[{{Tsygankov} {et~al.}(2019{\natexlab{a}}){Tsygankov}, {Doroshenko},
  {Mushtukov}, {Suleimanov}, {Lutovinov}, \& {Poutanen}}]{Tsygankov2019A05}
{Tsygankov}, S.~S., {Doroshenko}, V., {Mushtukov}, A.~A., {et~al.}
  2019{\natexlab{a}}, \mnras, 487, L30

\bibitem[{{Tsygankov} {et~al.}(2006){Tsygankov}, {Lutovinov}, {Churazov}, \&
  {Sunyaev}}]{Tsygankov2006}
{Tsygankov}, S.~S., {Lutovinov}, A.~A., {Churazov}, E.~M., \& {Sunyaev}, R.~A.
  2006, \mnras, 371, 19

\bibitem[{{Tsygankov} {et~al.}(2019{\natexlab{b}}){Tsygankov}, {Rouco
  Escorial}, {Suleimanov}, {Mushtukov}, {Doroshenko}, {Lutovinov}, {Wijnand s},
  \& {Poutanen}}]{Tsygankov2019-GX}
{Tsygankov}, S.~S., {Rouco Escorial}, A., {Suleimanov}, V.~F., {et~al.}
  2019{\natexlab{b}}, \mnras, 483, L144

\bibitem[{{Tuo} {et~al.}(2020){Tuo}, {Ji}, {Tsygankov}, {Mihara}, {Song}, {Ge},
  {Nabizadeh}, {Tao}, {Qu}, {Zhang}, {Zhang}, {Zhang}, {Bu}, {Chen}, {Xu},
  {Cao}, {Chen}, {Liu}, {Cai}, {Chang}, {Chen}, {Chen}, {Chen}, {Chen}, {Cui},
  {Cui}, {Deng}, {Dong}, {Du}, {Fu}, {Gao}, {Gao}, {Gao}, {Gu}, {Guan}, {Guo},
  {Han}, {Huang}, {Huo}, {Jia}, {Jiang}, {Jiang}, {Jin}, {Jin}, {Kong}, {Li},
  {Li}, {Li}, {Li}, {Li}, {Li}, {Li}, {Li}, {Li}, {Li}, {Li}, {Liang}, {Liao},
  {Liu}, {Liu}, {Liu}, {Liu}, {Liu}, {Lu}, {Lu}, {Lu}, {Luo}, {Luo}, {Ma},
  {Meng}, {Nang}, {Nie}, {Ou}, {Sai}, {Shang}, {Song}, {Sun}, {Tan}, {Wang},
  {Wang}, {Wang}, {Wang}, {Wang}, {Wen}, {Wu}, {Wu}, {Wu}, {Xiao}, {Xiao},
  {Xiong}, {Yang}, {Yang}, {Yang}, {Yang}, {Yi}, {Yin}, {You}, {Zhang},
  {Zhang}, {Zhang}, {Zhang}, {Zhang}, {Zhang}, {Zhang}, {Zhang}, {Zhang},
  {Zhang}, {Zhang}, {Zhang}, {Zhang}, {Zhang}, {Zhang}, {Zhang}, {Zhang},
  {Zhao}, {Zhao}, {Zheng}, {Zheng}, {Zhou}, {Zhou}, {Zhu}, {Zhu}, \&
  {Zhuang}}]{Tuo2020}
{Tuo}, Y.~L., {Ji}, L., {Tsygankov}, S.~S., {et~al.} 2020, Journal of High
  Energy Astrophysics, 27, 38

\bibitem[{{Verner} {et~al.}(1996){Verner}, {Ferland}, {Korista}, \&
  {Yakovlev}}]{Verner1996}
{Verner}, D.~A., {Ferland}, G.~J., {Korista}, K.~T., \& {Yakovlev}, D.~G. 1996,
  \apj, 465, 487

\bibitem[{{West} {et~al.}(2017){West}, {Wolfram}, \&
  {Becker}}]{2017ApJ...835..130W}
{West}, B.~F., {Wolfram}, K.~D., \& {Becker}, P.~A. 2017, \apj, 835, 130

\bibitem[{{Willingale} {et~al.}(2013){Willingale}, {Starling}, {Beardmore},
  {Tanvir}, \& {O'Brien}}]{Willingale2013}
{Willingale}, R., {Starling}, R.~L.~C., {Beardmore}, A.~P., {Tanvir}, N.~R., \&
  {O'Brien}, P.~T. 2013, \mnras, 431, 394

\bibitem[{{Wilms} {et~al.}(2000){Wilms}, {Allen}, \& {McCray}}]{Wilms2000}
{Wilms}, J., {Allen}, A., \& {McCray}, R. 2000, \apj, 542, 914

\bibitem[{{Xiao} {et~al.}(2019){Xiao}, {Ji}, {Staubert}, {Ge}, {Zhang},
  {Zhang}, {Santangelo}, {Ducci}, {Liao}, {Guo}, {Li}, {Zhang}, {Qu}, {Lu},
  {Li}, {Song}, {Xu}, {Bu}, {Cai}, {Cao}, {Chang}, {Chen}, {Chen}, {Chen},
  {Chen}, {Chen}, {Chen}, {Cui}, {Cui}, {Deng}, {Dong}, {Du}, {Fu}, {Gao},
  {Gao}, {Gao}, {Gu}, {Guan}, {Gungor}, {Guo}, {Han}, {Huang}, {Huo}, {Jia},
  {Jiang}, {Jiang}, {Jin}, {Kong}, {Li}, {Li}, {Li}, {Li}, {Li}, {Li}, {Li},
  {Li}, {Li}, {Liang}, {Liu}, {Liu}, {Liu}, {Liu}, {Liu}, {Lu}, {Lu}, {Luo},
  {Luo}, {Ma}, {Meng}, {Nang}, {Nie}, {Ou}, {Sai}, {Song}, {Sun}, {Tan}, {Tao},
  {Tuo}, {Wang}, {Wang}, {Wang}, {Wang}, {Wang}, {Wen}, {Wu}, {Wu}, {Wu},
  {Xiong}, {Yang}, {Yang}, {Yang}, {Yang}, {Yin}, {Yin}, {Zhang}, {Zhang},
  {Zhang}, {Zhang}, {Zhang}, {Zhang}, {Zhang}, {Zhang}, {Zhang}, {Zhang},
  {Zhang}, {Zhang}, {Zhang}, {Zhang}, {Zhang}, {Zhang}, {Zhao}, {Zhao},
  {Zheng}, {Zhou}, {Zhu}, \& {Zhu}}]{NuSTAR-HXMT-Calibration2019}
{Xiao}, G.~C., {Ji}, L., {Staubert}, R., {et~al.} 2019, Journal of High Energy
  Astrophysics, 23, 29

\bibitem[{{Zhang} {et~al.}(2020){Zhang}, {Li}, {Lu}, {Song}, {Xu}, {Liu},
  {Chen}, {Cao}, {Bu}, {Chang}, {Chen}, {Chen}, {Chen}, {Chen}, {Chen}, {Cui},
  {Cui}, {Deng}, {Dong}, {Du}, {Fu}, {Gao}, {Gao}, {Gao}, {Ge}, {Gu}, {Guan},
  {Gungor}, {Guo}, {Han}, {Hu}, {Huang}, {Huo}, {Jia}, {Jiang}, {Jiang}, {Jin},
  {Jin}, {Li}, {Li}, {Li}, {Li}, {Li}, {Li}, {Li}, {Li}, {Li}, {Li}, {Li},
  {Liang}, {Liao}, {Liu}, {Liu}, {Liu}, {Liu}, {Liu}, {Liu}, {Lu}, {Lu}, {Luo},
  {Ma}, {Meng}, {Nang}, {Nie}, {Ou}, {Qu}, {Sai}, {Shang}, {Shen}, {Sun},
  {Tan}, {Tao}, {Tuo}, {Wang}, {Wang}, {Wang}, {Wang}, {Wang}, {Wang}, {Wang},
  {Wen}, {Wu}, {Wu}, {Wu}, {Xiao}, {Xiong}, {Yan}, {Yang}, {Yang}, {Yang},
  {Yi}, {Yuan}, {Zhang}, {Zhang}, {Zhang}, {Zhang}, {Zhang}, {Zhang}, {Zhang},
  {Zhang}, {Zhang}, {Zhang}, {Zhang}, {Zhang}, {Zhang}, {Zhang}, {Zhang},
  {Zhang}, {Zhang}, {Zhang}, {Zhang}, {Zhang}, {Zhao}, {Zhao}, {Zheng}, {Zhou},
  {Zhu}, {Zhu}, {Zhuang}, \& {Insight-HXMT team}}]{Zhang2020}
{Zhang}, S.-N., {Li}, T., {Lu}, F., {et~al.} 2020, Science China Physics,
  Mechanics, and Astronomy, 63, 249502

\end{thebibliography}

\end{document}